\documentclass[twocolumn, pre,aps,10pt]{revtex4-1}

\usepackage{amsfonts,amssymb}
\usepackage[sumlimits,intlimits]{amsmath}
\usepackage{graphics}
\usepackage{graphicx}
\usepackage{epsfig}
\usepackage{mathrsfs}
\usepackage{verbatim}
\usepackage{hyperref}    
\usepackage{bm}          

\newcommand{\be}{\begin{equation}}
\newcommand{\ee}{\end{equation}}
\newcommand{\ki}{\kappa_\text{i}}
\newcommand{\knc}{\kappa_\text{NC}}
\newcommand{\Nox}{N_\text{imp}}

\begin{document}

\title{Theory of hopping conduction in arrays of doped semiconductor nanocrystals}

\date{\today}

\author{Brian Skinner}
\author{Tianran Chen}
\author{B. I. Shklovskii}
\affiliation{Fine Theoretical Physics Institute, University of Minnesota, Minneapolis, MN 55455, USA}

\begin{abstract}

The resistivity of a dense crystalline array of semiconductor nanocrystals (NCs) depends in a sensitive way on the level of doping as well as on the NC size and spacing.  The choice of these parameters determines whether electron conduction through the array will be characterized by activated nearest-neighbor hopping or variable-range hopping (VRH).  Thus far, no general theory exists to explain how these different behaviors arise at different doping levels and for different types of NCs.  In this paper we examine a simple theoretical model of an array of doped semiconductor NCs that can explain the transition from activated transport to VRH.  We show that in sufficiently small NCs, the fluctuations in donor number from one NC to another provide sufficient disorder to produce charging of some NCs, as electrons are driven to vacate higher shells of the quantum confinement energy spectrum.  This confinement-driven charging produces a disordered Coulomb landscape throughout the array and leads to VRH at low temperature.  We use a simple computer simulation to identify different regimes of conduction in the space of temperature, doping level, and NC diameter.  We also discuss the implications of our results for large NCs with external impurity charges and for NCs that are gated electrochemically.

\end{abstract}
\maketitle

\section{Introduction} \label{sec:intro}

Arrays of semiconductor nanocrystals (NCs) have great promise for optoelectronic and photovoltaic devices.  The usefulness of NC arrays comes from the ability to tune both their optical properties -- generally by choosing the size or shape of NCs \cite{Jurbergs2006snw, Rafiq2006hci} -- and their electronic properties -- usually through the addition of dopants or surface ligands that control the spacing between NCs \cite{Moreira2011ect, Yu2003ncc}.  Recent experiments have demonstrated that dense, crystalline arrays of spherical semiconductor NCs can be reliably produced with diameter in the range $4$--$10$ nm and with less than 5\% dispersion \cite{Jurbergs2006snw, Talapin2010poc}.  Thus, optoelectronic or photovoltaic devices made from NCs can be designed to operate precisely in any chosen region of the optical spectrum.

From a practical standpoint, however, the development of NC-based devices is slowed by the high resistivity of the NC arrays.  In their undoped state, semiconductor NCs are insulators, and in order to reduce their large resistivity it is necessary to bring additional electrons (or holes) to the NCs either through chemical doping \cite{Norris2008dn} or electrochemical gating \cite{Liu2010mae}.  In this article we focus primarily on the former, although we comment on electrochemical gating at the end of the paper.

\begin{figure}[tb!]
\centering
\includegraphics[width=0.4 \textwidth]{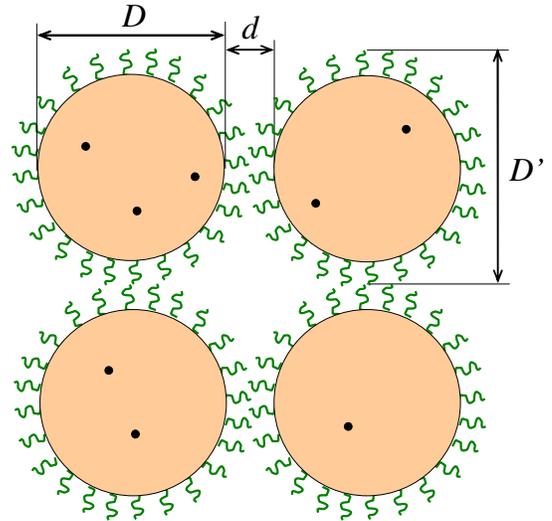}
\caption{(Color online) Schematic drawing of spherical semiconductor NCs (large, light-colored circles) with diameter $D$ arranged in a crystalline lattice with lattice constant $D'$.  Each NC is coated in a thin layer of insulating ligands (curvy lines) that maintain a separation $d = D' - D$ between NCs and prevent them from sintering.  Each NC has a random number of donors in its interior (small, black circles).} \label{fig:schematic}
\end{figure}

In particular, we consider the case where each NC is made from a semiconductor that is heavily-doped, for example, by donor impurities.  In this case all donor electrons reside in the conduction band of the NC.  In order to conduct across the array, these electrons must tunnel between NCs under the high barrier associated with the insulator (such as the ligands shown in Fig.\ \ref{fig:schematic}) that fills the space between them.  

In the presence of even a relatively small amount of disorder in the array, the large tunneling barriers imply that donor electrons experience Anderson localization due to fluctuations in the electron energy from one NC to another \cite{Mott1972itc}.  In this situation conduction proceeds only by phonon-assisted tunneling, or ``hopping", between localized electron states.  This hopping is a thermally-activated process in which electron tunneling occurs simultaneously with the absorption or emission of a phonon whose energy accounts for the difference between the initial and final electron states.  
(While metallic conduction through the array is in principle possible, and has been reported \cite{Kagan2012ros}, it requires the characteristic disorder energy in the system to be smaller than the hopping integral $t$ between neighboring NCs.  Since $t$ decays exponentially with the separation $d$ between NCs and with the height of the tunneling barrier them, the condition for metallic conductivity is difficult to meet, and in this paper we assume that electron conduction proceeds by hopping.)

If one assumes that in the global ground state of the array all NCs are neutral, then hopping transport requires an electron to be thermally excited to jump from one neutral NC to another.  This process produces two oppositely charged NCs, each of which has a corresponding Coulomb self-energy $\varepsilon_c = e^2/\kappa D$, where $\kappa$ is the effective dielectric constant of the NC array and $D$ is the NC diameter.  This charging energy plays the role of an activation energy for resistivity in the case where all NCs are neutral in the global ground state.  Equivalently, one can say that the distribution of electron ground state energies, or the ``density of ground states" (DOGS) of NCs, has a gap of width $2 \varepsilon_c$ centered at the electron Fermi level.  As a result, the resistivity $\rho$ follows the Arrhenius law: $\ln \rho \propto \varepsilon_c/k_BT$, where $k_BT$ is the thermal energy.  We emphasize that the activation energy for hopping conduction is sensitive only to the \emph{ground state} energies of electrons and holes that are added to NCs.  For this reason when calculating the resistivity it is sufficient to consider the DOGS, which does not include excited electron states with additional kinetic energy.

In experiments, however, one often observes a temperature dependence of the resistivity that is different from simple activation: $\ln \rho \propto T^{-\gamma}$, with the temperature exponent $\gamma < 1$.  Such ``stretched exponential" behavior is believed to be possible only if the disorder is so strong that a substantial fraction of NCs is charged in the global ground state.  Such charging creates a random Coulomb potential landscape that shifts up and down the electron energy spectra at different NCs.  As a result of this shifting, the gap in the DOGS is smeared and filled.  This smearing means that some electron states have energies very close to the Fermi level, and as a result one can find a pair of empty and filled electron states separated by an energy $\Delta \varepsilon$ that is much smaller than $\varepsilon_c$.  At small temperature $k_BT \ll \varepsilon_c$, it is hopping between such pairs that are close in energy that dominates the conduction.  

Of course, for small $\Delta \varepsilon$ the typical separation $r$ between the corresponding NC pair is much larger than the spacing $D'$ between neighboring NCs.  Thus, at small temperature $T$ electron conduction relies on tunneling between distant NCs.  To understand how such long-range tunneling is possible, consider first the tunneling of an electron between nearest-neighboring NCs.  When the electron tunnels through the insulating gap of thickness $d$ between NCs, it accumulates an action $\hbar d/a$, where $a$ is the decay length of the electron wavefunction outside of the NC.  Thus, the tunneling amplitude between nearest neighbors is suppressed by a factor $\sim \exp[-d/a]$.  On the other hand, when an electron tunnels to a NC at a distance $x \gg D'$, the path of least action for the electron is to travel primarily through nearest-neighboring NCs, making hops only through the small gaps between neighbors and thereby accumulating an action $\sim \hbar (d/a)(x/D')$, plus an additional much smaller term corresponding to action accumulated across the interior of each NC.  Thus, the tunneling amplitude to the distance $x$ is suppressed by a factor $\sim \exp[-x d/D'a]$.  The exponential decay of the tunneling amplitude is described by defining the localization length $\xi$, such that tunneling between NCs with separation $r$ is suppressed by the factor $\exp[-2 r/\xi]$.  By the argument above, one cannot simply equate $\xi$ with $a$, but rather $\xi \sim a D'/d \gg a$ \cite{Zhang2004dos}.  It is this enhanced localization length, made possible by tunneling through intermediate NCs, that allows for long-range hopping.  In the remainder of this paper, we consider the limit where $d$ and $a$ are both very small compared to the NC diameter, so that $D' \simeq D$ while $\xi$ remains finite.   

If the temperature $T$ is made increasingly small, the corresponding energy difference $\Delta \varepsilon$ of electron hops becomes increasingly small due to the scarcity of available high-energy phonons, and as a result the typical hop length increases.  Such behavior is known as variable range hopping (VRH), and is responsible for the stretched exponential behavior $\gamma < 1$ in the resistivity.  When the DOGS is constant near the Fermi level, the resistivity follows the Mott law of VRH \cite{Mott1968cig}: $\ln \rho \propto T^{-1/4}$.  However, in systems where the long-ranged Coulomb potential is not screened, electron correlation effects produce a DOGS that vanishes quadratically with energy at the Fermi level \cite{Efros1975cga}.  Such a vanishing DOGS results in the Efros-Shklovskii (ES) law of VRH: $\ln \rho \propto T^{-1/2}$.  
In principle, all three of these conduction behaviors --- Arrhenius ($\gamma = 1$), Mott VRH ($\gamma = 1/4$), and ES VRH ($\gamma = 1/2$) --- are possible in arrays of semiconductor NCs, depending on the magnitude and type of disorder present.  In this paper we focus our description on the fundamental role played by inherent fluctuations in donor number among doped NCs.

Experiments probing the resistivity of NC arrays have reported that the resistivity depends in a sensitive and qualitative way on the level of doping \cite{Liu2010mae}.  Specifically, as the average number $\nu$ of dopant electrons per NC is varied, the dependence of the resistivity $\rho$ on the temperature $T$ changes between Arrhenius-type activated conduction ($\gamma = 1$) and VRH ($\gamma < 1$).  VRH has been reported in a variety of granular semiconductor systems \cite{Liu2010mae, Rafiq2006hci, Romero2005cba, Wienkes2012eti}, but thus far there is no general theory to explain how these different types of conduction can coexist and why they appear in particular ranges of the electron ``filling factor" $\nu$.

In this paper we present such a theory, based on a first-principles description of the ground state arrangement of electrons within an array of doped NCs.  We focus on a simple model of identical spherical NCs that are covered by a thin layer of insulating ligand (or some other insulator) and  arranged in an ideal crystalline lattice, as depicted in Fig.\ \ref{fig:schematic}.  We show that the presence of fluctuations in donor number between different NCs is sufficient to produce charging of NCs, which results in a disordered Coulomb landscape that encourages VRH.  This charging is driven by the large gaps between shells of the electron quantum energy spectrum in NCs with large Bohr radius $a_B$.  Specifically, these inter-shell gaps drive electrons to depart from NCs with a large number of donors, where maintaining electroneutrality would require placing electrons in higher quantum energy shells, and reside instead on nearby NCs with small donor number.  In this way some NCs spontaneously acquire a positive or negative charge, and it is this charging that leads to VRH when the temperature is not too large.

Using this model, we explain how the different regimes of resistivity observed in experiment arise based on the interplay between the charging spectrum of NCs, the long-ranged Coulomb interactions between charged NCs, and the discrete quantum energy levels of confined electrons.  We supplement our theory with a simple computer simulation, which we use to calculate the DOGS and the resistivity.  

\begin{figure}[htb!]
\centering
\includegraphics[width=0.5 \textwidth]{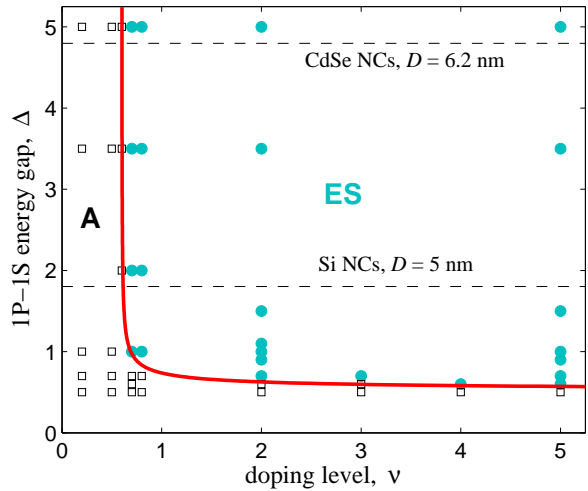}
\caption{(Color online) Phase diagram indicating regimes of activated and ES resistivity as a function of doping level $\nu$ and the dimensionless quantum energy gap $\Delta \equiv 20.64 \kappa a_B/\knc D$ at low temperature $k_BT \ll e^2 \xi/\kappa D^2$.  Symbols correspond to simulated systems: filled (light blue) circles indicate systems that exhibited ES resistivity and open squares indicate systems that exhibited activated resistivity.  The simulation method is described in detail in Sec.\ \ref{sec:computer}.  The thick (red) curve is an approximate boundary between these two regimes, which are labeled ``ES" and ``A", respectively.  Dashed, horizontal lines indicate the value of $\Delta$ corresponding to Si NCs with $D = 5$ nm (as in Ref.\ \cite{Jurbergs2006snw}) and to CdSe NCs with $D = 6.2$ nm (as in Ref.\ \cite{Liu2010mae}).  This phase diagram is discussed more thoroughly in Sec.\ \ref{sec:results}.} \label{fig:phasediagram}
\end{figure}

Our main result is that VRH appears when the average number $\nu$ of electrons per NC, the NC diameter $D$, and the temperature $T$ satisfy the following three conditions:
\begin{itemize}
 \item[(i)] $\nu \gtrsim 0.6$, 
\item[(ii)] $D \lesssim 34 \kappa a_B/\knc$, and 
\item[(iii)] $k_BT \lesssim 0.5 e^2 \xi/\kappa D^2$.  
\end{itemize}
Here, $\knc$ is the internal dielectric constant of NCs.  When these three conditions are satisfied, the resistivity follows the ES law.  In situations where any of the three criteria is not met, the conduction is activated.  This result is depicted at low temperature, $k_BT \ll e^2 \xi/\kappa D^2$, in the phase diagram of Fig.\ \ref{fig:phasediagram}.

The remainder of this paper is organized as follows.  In Sec.\ \ref{sec:model} we define the theoretical model to be studied.  Sec.\ \ref{sec:computer} describes our computer simulation, including our methods for numerically calculating the DOGS and resistivity.  Results are presented in Sec.\ \ref{sec:results}, along with a discussion of why Arrhenius and VRH resistivity appear in particular regimes of $\nu$, $D$, and $T$.  We also discuss interesting features of the DOGS in this model, including the appearance of ``reflected Coulomb gaps" at either side of the Fermi level.

In Sec.\ \ref{sec:largeoxide} we discuss the implications of our theory for large NCs with external impurity charges.  We present a modified model appropriate for this case and we arrive at a single condition for VRH associated with the concentration of external impurity charges.  Sec.\ \ref{sec:IL} presents some speculation on how our results can be applied to electrochemical gating of NC arrays using ionic liquids, and this is followed by concluding remarks in Sec.\ \ref{sec:conclusion}.

\section{Model of NC arrays with random number of dopants} \label{sec:model}

In this paper our goal is to describe the resistivity of a dense array of semiconductor NCs and capture its dependence on doping level, temperature, and NC diameter.  To this end we adopt the following simplified theoretical model.  We consider NCs to be identical spheres of diameter $D$ with large internal dielectric constant $\knc \gg \kappa$.  These spheres are arranged in a regular, three-dimensional (3D) lattice, with each lattice site $i$ located at the center of a NC.  For simplicity, we consider a cubic lattice with lattice constant $D'$ just barely larger than $D$, so that $d \ll D$ (see Fig.\ \ref{fig:schematic}).  Our choice of a cubic lattice does not qualitatively affect any of the results we present below. 

We further assume that the radius $D/2$ of the NCs is comparable to or smaller than the effective electron Bohr radius $a_B = \hbar^2 \knc/m e^2$ of the semiconductor, where $e$ is the electron charge and $m$ is the effective electron mass.  As an example, NCs made from Si have $a_B \approx 2.4$ nm; for CdSe NCs, $a_B \approx 5$ nm.  Under this condition the wavefunction of a donor electron is extended across the entire volume of a NC, rather than localized around a donor impurity, and the energy of the electron is strongly affected by quantum confinement within the NC.  As an example, a single donor in the center of a NC has a delocalized electron state when $D < 6 a_B$ \cite{Ekimov1990sad, Efros2000eso}.  This condition can be used as a somewhat conservative estimate for how small the diameter should be to produce electron states that are extended across the NC.

In order to obtain the quantum energy spectrum in NCs, one can make the approximation that each NC is an infinite 3D square well.  Such an approximation is valid because of the NCs' relatively large work function.  The resulting energy spectrum can be described by defining the energy $E_Q(n)$ of the $n$th lowest electron, which gives for the first few energy levels
\be 
E_Q(n) = \frac{\hbar^2}{m D^2} \times \left\{ 
\begin{array}{lr}
0, &  n = 0 \\
19.74, &  n = 1, 2 \\
40.38, &  3 \leq n \leq 8 \\
66.43, &  9 \leq n \leq 18
\end{array}
\right.
\label{eq:EQ}.
\ee
These first three nonzero energy levels can be labeled 1S, 1P, and 1D, respectively.  Higher electron shells have thus far not been examined by experiment, since they correspond to very large doping, and will not be discussed in this work.  We focus primarily on the case where the spacing between quantum energy levels $\approx 20 \hbar^2/mD^2$ is larger than the characteristic scale of Coulomb energies, $e^2/\kappa D$.  The expression of Eq.\ (\ref{eq:EQ}) ignores the weak perturbation of quantum energy levels resulting from electron-electron interactions.  This approximation is justified because of the large internal dielectric constant $\knc$, as explained below.

During the doping process, each NC $i$ acquires some number $N_i$ of positively-charged donors that it contains within its interior.  These are assumed to be fixed, while the number of electrons $n_i$ within the NC can change due to electron tunneling between NCs.  We assume that donors are added randomly to each NC by some high-temperature process, so that if the average number of donors per NC is $\nu$, then the probability that a given NC will have exactly $N$ donors is given by the Poisson distribution:
\be 
P(N) = \frac{\nu^N}{N!} e^{-\nu}.
\label{eq:Poisson}
\ee
This randomness in the number of donors is the only form of disorder that we include in our model.  We show in Sec.\ \ref{sec:results} that this disorder is sufficient to produce random charging of NCs, which leads to VRH.  As mentioned in the introduction, the spontaneous charging of NCs is the result of the large gaps between quantum kinetic energy shells, which drive electrons away from NCs with many donors (emptying higher shells) and into NCs with few donors (filling lower shells), so that the number of electrons in a given NC is not generally equal to the number of donors.
Additional disorder arising from fluctuations in the NC size is not considered explicitly in this paper.  The possible effect of such size fluctuations is discussed at the end of Sec.\ \ref{sec:results}, but we note here that fluctuation of NC size alone cannot produce spontaneous charging of NCs in the global ground state, which, as we show below, plays a crucial role for VRH.

In addition to the quantum kinetic energy of the system, transport through the array is also greatly affected by long-ranged Coulomb interactions, which must be taken into account.  In general, one could expect that calculating the total Coulomb energy of the system is a difficult problem, since the positions of negative electrons within each NC are described by their corresponding quantum wavefunctions and the positions of positive donors are random within the NC's volume.  For our problem, however, a significant simplification is available because the internal dielectric constant $\knc$ is much larger than both the external dielectric constant $\ki$ of the insulator in which the NCs are embedded and the overall effective dielectric constant $\kappa$ of the assembly.  Specifically, the large internal dielectric constant $\knc$ implies that any internal charge $e$ is essentially completely compensated by the dielectric response, with the great majority of that charge, $e (\knc - \kappa)/\knc$, becoming distributing across the surface of the NC.  In this way each NC can be thought of as metallic in terms of its Coulomb interactions.  This allows us to write that the Coulomb self-energy of a NC with net charge $q$ is given approximately by $q^2/\kappa D$, irrespective of how its constituent internal charges are arranged.  The interaction between two NCs $i,j$ at a distance $r_{ij}$ can also be approximated as $q_i q_j/\kappa r_{ij}$.  These approximations are equivalent to the so-called constant interaction model, which is commonly used for individual quantum dots \cite{Meir1991tts}.

It should be noted that the effective dielectric constant $\kappa$ of the NC array is not simply equal to the dielectric constant  $\ki$ of the insulating medium between NCs, but also includes the effect of polarization of NCs in response to an applied field.  This polarization effectively decreases both the Coulomb self-energy of a single NC and the interaction between neighboring NCs.  Generally speaking, the renormalization of the dielectric constant is not very strong, so that $\kappa$ is not very different from $\ki$ even when $\knc \gg \ki$.  The canonical Maxwell-Garnett formula gives the approximate relation \cite{Maxwell1891tea}
\be 
\kappa \simeq \ki \frac{\knc + 2 \ki + 2 f(\knc - \ki)}{\knc + 2 \ki - f (\knc - \ki)},
\label{eq:Maxwell}
\ee
where $f = \pi D^3/[6 (D')^3]$ is the volume fraction occupied by the NCs; for $f < 0.4$, this expression is accurate to within 8\% \cite{Doyle1978cpc}.  As an example, for the case of a cubic lattice with $D = 5$ nm and $D' = 6$ nm (so that $f = 0.3$) and for $\knc/\ki = 5$, one has $\kappa \approx 1.6 \ki$.

Given this model, we can write down the Hamiltonian for our system as
\begin{eqnarray}
H & = & \sum_i \left[ \frac{e^2 (N_i - n_i)^2 }{\kappa D} + \sum_{k = 0}^{n_i}E_Q(k) \right] \nonumber \\
& & + \sum_{\langle i,j \rangle } \frac{e^2 (N_i - n_i)(N_j - n_j)}{\kappa r_{ij}}.
\label{eq:H}
\end{eqnarray}
Here, the first term describes the electrostatic self-energy of NC $i$, which has charge $q_i = e(N_i - n_i)$, the second term describes the total quantum energy of the $n_i$ electrons on NC $i$, and the last term indicates the Coulomb interaction between different NCs.

The ground state for a particular system (a set of donor numbers $\{N_i\}$) is defined by the set of electron occupation numbers $\{n_i\}$ that minimizes the Hamiltonian $H$.  Given the ground state configuration, one can determine the energy of the highest filled electron level, $\varepsilon_i^{(f)}$, and the lowest empty electron level, $\varepsilon_i^{(e)}$, at each NC $i$.  Specifically,
\begin{eqnarray} 
\varepsilon_i^{(f)} & = & E_Q(n_i) + \frac{e^2[(N_i - n_i)^2 - (N_i - n_i + 1)^2]}{\kappa D} \nonumber \\
& & - \sum_{j \neq i} \frac{e(N_j - n_j)}{\kappa r_{ij}}
\label{eq:enf}
\end{eqnarray}
and
\begin{eqnarray} 
\varepsilon_i^{(e)} & = & E_Q(n_i + 1) + \frac{e^2[(N_i - n_i - 1)^2 - (N_i - n_i)^2]}{\kappa D} \nonumber \\
& & - \sum_{j \neq i} \frac{e(N_j - n_j)}{\kappa r_{ij}}.
\label{eq:ene}
\end{eqnarray}
For the global ground state configuration, $\varepsilon_i^{(f)} < \varepsilon_j^{(e)}$ for all $i, j$.  As alluded to in the introduction, the definitions of $\varepsilon_i^{(f)}$ and $\varepsilon_i^{(e)}$ describe only the lowest energy state of an electron or hole added to the site $i$.  For this reason we refer to the density of states of these energy states $\varepsilon_i^{(e,f)}$ as the DOGS.

The resistivity of the NC array is largely determined by the set of these ground state single-particle energies $\{\varepsilon_i^{(f)} \}$ and $\{\varepsilon_i^{(e)}\}$.  In the following section we show how these energy states can be used to calculate both the ground state electron DOGS $g(\varepsilon)$ and the resistivity $\rho$ as a function of temperature and doping level.  Note that in this problem every site is represented by two energies, in contrast to the canonical impurity band of lightly-doped semiconductors \cite{Efros1984epo}, where every donor has only one relevant excitation energy.

It is also important to note that in our model these donor electrons are assumed to be responsible for all conduction.  In other words, we assume that the temperature $T$ is low enough (and the doping level $\nu$ is high enough) that donor electrons are much more abundant than electrons activated from the valence band.  In practical cases, this assumption is easily met: it requires only that the thermal energy $k_BT$ be much smaller than the band gap energy $E_g$.  More exactly, it requires that $k_BT \ll E_g/\ln [\knc D^2 E_g/e^2 a_B \nu^{2/3}]$.

\section{Computer modeling} \label{sec:computer}

In this section we describe our computational method for calculating the density of states and the resistivity at a given value of $\nu$, $T$, and $D$.  These calculations are based on a computer simulation of a finite, cubic array of $L \times L \times L$ NCs, which proceeds as follows.  First, we specify the doping level $\nu$.  The simulation then assigns the donor number $N_i$ for each NC $i$ according to Eq.\ (\ref{eq:Poisson}).  The initial values of the electron numbers $\{n_i\}$ are then assigned randomly in such a way that the system is overall electro-neutral, i.e., $\sum_i n_i = \sum_j N_j$.  The simulation then searches for the ground state by looping over all NC pairs $\langle i j \rangle$ and attempting to move one electron from $i$ to $j$.  If the move lowers the Hamiltonian $H$, then it is accepted, otherwise it is rejected.  Equivalently, one can say that for each pair $i, j$ we check that two ES ground state criteria are satisfied:
\be 
\varepsilon_j^{(e)} - \varepsilon_{i}^{(f)} - \frac{e^2}{\kappa r_{ij}} > 0
\label{eq:EScrit1}
\ee 
and
\be 
\varepsilon_i^{(e)} - \varepsilon_j^{(f)} - \frac{e^2}{\kappa r_{ij}} > 0.
\label{eq:EScrit2}
\ee
If either one of these criteria is violated, then an electron is transferred.  This process continues until all sites $i,j$ satisfy Eqs.\ (\ref{eq:EScrit1}) and (\ref{eq:EScrit2}).

It should be noted that this procedure does not in general find the exact ground state, but only a ``pseudo-ground state" that is stable with respect to single-electron transfers.  In principle, the system energy can be lowered further by some multi-electron transfers.  The effect of these higher-order relaxation processes on the properties of the pseudo-ground state has been examined for similar models \cite{Mobius1992cgi, Efros2011cgi}, and they are generally beyond our intended accuracy in this paper, so we do not consider them here.

Once the pseudo-ground state occupation numbers $\{ n_i \}$ have been found, one can define the single-particle energies $\varepsilon_i^{(f)}$ and $\varepsilon_i^{(e)}$ for each NC $i$ using Eqs.\ (\ref{eq:enf}) and (\ref{eq:ene}).  These energies are tabulated and then histogrammed in order to calculate the single-particle DOGS $g(\varepsilon)$.  In the results presented below we define electron energies $\varepsilon$ relative to the Fermi level $\mu$, which is calculated for each realization of the simulation as $\mu = [ \min\{ \varepsilon_i^{(e)} \} - \max\{ \varepsilon_i^{(f)} \}]/2$.  In this way $\varepsilon < 0$ corresponds to filled electron states $\varepsilon^{(f)}$ while $\varepsilon > 0$ corresponds to empty states $\varepsilon^{(e)}$.  (See, for example, Fig.\ \ref{fig:nu5} below.)

Once the pseudo-ground state energies $\{\varepsilon_i^{(f)} \}$ and $\{\varepsilon_i^{(e)}\}$ are determined, we calculate the resistivity of the system by mapping the simulated NC array to an effective resistor network.  The equivalent resistance $R_{ij}$ between NCs $i$ and $j$ can be determined by writing down the time-averaged rate of electron transfer between sites $i$ and $j$ in the presence of an electric field and expanding in the limit of small field, as in the canonical Miller-Abrahams resistor network \cite{Efros1984epo, Miller1960ica}.  In calculating $R_{ij}$ we consider only electron transfer among the highest filled states, $\varepsilon^{(f)}$, and the lowest empty states, $\varepsilon^{(e)}$, which is appropriate when the temperature is small enough that $T < e^2/\kappa D$, so that thermal excitation of multi-electron transitions is exponentially unlikely.

Since each NC has two energy levels that can participate in conduction, $\varepsilon^{(f)}$ and $\varepsilon^{(e)}$, one can say that there are four parallel conduction processes that contribute to the resistivity between two NCs $i$ and $j$: one for each combination of the initial energy level at site $i$ (either $\varepsilon_i^{(f)}$ or $\varepsilon_i^{(e)}$) and the final energy level at site $j$ (either $\varepsilon_j^{(f)}$ or $\varepsilon_j^{(e)}$).  Each of these four processes has a corresponding effective resistance $R_{ij}^{(\alpha \beta)}$, where $\alpha, \beta = (f), (e)$.  These four resistances can be said to be connected in parallel between NCs $i$ and $j$, and their value can be written compactly as
\be 
R_{ij}^{(\alpha \beta)} = R_0 \exp\left[\frac{2 r_{ij}}{\xi} + \frac{\varepsilon_{ij}^{(\alpha, \beta)}}{k_B T} \right],
\label{eq:Rijab}
\ee 
where $R_0$ is a prefactor that has only a relatively weak power-law dependence on temperature.  The first term in the exponential of Eq.\ (\ref{eq:Rijab}) describes the exponential suppression of the tunneling rate with distance $r$, as explained in the introduction, and the second term describes thermal activation by exponentially-rare phonons of energy $\varepsilon_{ij}^{(\alpha, \beta)}$.  Since we are interested only in identifying the exponential component of the dependence of resistivity on temperature, we take $R_0$ to be a constant.  The energy $\varepsilon_{ij}^{(\alpha, \beta)}$ in Eq.\ (\ref{eq:Rijab}) is defined as follows \cite{Efros1984epo}:
\be 
\varepsilon_{ij}^{(\alpha, \beta)} = \left\{
\begin{array}{lr}
|\varepsilon_{j}^{(\beta)} - \varepsilon_{i}^{(\alpha)}| - \frac{e^2}{\kappa r_{ij}}, &  \varepsilon_{j}^{(\beta)}\varepsilon_{i}^{(\alpha)} < 0 \vspace{2mm} \\
\max \left[ \left|\varepsilon_{j}^{(\alpha)} \right|, \left|\varepsilon_{j}^{(\beta)} \right| \right], &  \varepsilon_{j}^{(\beta)}\varepsilon_{i}^{(\alpha)} > 0
\end{array}
\right.
.
\label{eq:epsij}
\ee

The net resistance $R_{ij}$ between NCs $i$ and $j$ is the parallel sum of the four resistances $R_{ij}^{(\alpha \beta)}$.  Since the exponential factor in Eq.\ (\ref{eq:Rijab}) provides a sharp differentiation between these four parallel resistances, at relatively low temperatures and to within the accuracy of our calculations we can equate $R_{ij}$ with the minimum of the four parallel resistances.  That is,
\be 
R_{ij} \simeq \min \left\{ R_{ij}^{(\alpha \beta)} \right\}.
\ee
After calculating all resistances $R_{ij}$ for a given simulated array, we find the dimensionless resistivity of the network $\rho/\rho_0$, where $\rho_0 = R_0 D'$, using a percolation approach \cite{Efros1984epo}.  Specifically, we find the minimum value $R_c$ such that if all resistances $R_{ij}$ with $R_{ij} < R_c$ are left intact while others are eliminated (replaced by $R_{ij} = \infty$), then there exists a pathway connecting the left and right faces of the simulation volume (the ``infinite" percolation cluster).  The resistivity $\rho/\rho_0$ is approximated as $R_c/R_0$.


In our analysis below we make use of the following dimensionless units, which reduce the number of free variables in the problem.  We introduce the dimensionless distance between the centers of NCs $i$ and $j$,
\be 
r_{ij}^* = \frac{r_{ij}}{D},
\ee
the dimensionless temperature
\be 
T^* = \frac{2 D^2 \kappa k_B T}{e^2 \xi},
\label{eq:Tstar}
\ee 
the dimensionless electron energy
\be 
\varepsilon^* = \frac{\varepsilon}{e^2/\kappa D},
\label{eq:epsilonstar}
\ee 
the dimensionless electron DOGS
\be  
g^*(\varepsilon^*) = \frac{e^2 D^2}{\kappa}g(\varepsilon^*),
\ee
and the dimensionless resistivity
\be 
\ln \rho^* = \frac{\xi}{2 D} \ln (\rho/\rho_0).
\ee 
In these units, Eq.\ (\ref{eq:Rijab}) can be written more simply as
\be 
\ln \rho_{ij}^* = r_{ij}^* + \varepsilon_{ij}^*/T^*,
\label{eq:lnrho}
\ee 
and the problem loses any explicit dependence on the diameter or the localization length.  It is also convenient to discuss the energy gap between the 1S and 1P shells in terms of the dimensionless parameter
\be 
\Delta \equiv \frac{E_Q(3) - E_Q(2)}{e^2/\kappa D} = 20.64 \frac{\kappa \hbar^2}{m e^2 D} = 20.64 \frac{\kappa a_B}{\knc D}.
\label{eq:Deltadef}
\ee 
We use our simulation to examine the resistivity at various values of $\nu$, $T^*$, and $\Delta$.

Results below correspond to a simulated system of size $L = 25$ with open boundaries, averaged over 100 realizations.  Simulations at smaller system size, $15 \leq L < 25$, do not produce noticeably different results for either the DOGS or the resistivity, which allows us to avoid having to extrapolate our results to infinite system size.

\section{Results and discussion} \label{sec:results}

Our goal is to determine which conditions produce VRH in the NC array.  To this end we calculated the resistivity $\rho$ and the electron density of ground states $g(\varepsilon)$ for a range of values of the doping level $\nu$, the temperature $T^*$, and the quantum energy scale $\Delta$.  (Varying $\Delta$ is equivalent to considering different values of the NC diameter.)  Before proceeding to present general results, however, we first illustrate the most important features of the problem by discussing the hypothetical case where all NCs have the same number of donors, so that there is absolutely no disorder in the system.  Say, for example, that $\nu = 5$ and that $N_i = 5$ for all $i$.  In this situation, the ground state arrangement of the system is for electrons to uniformly neutralize all donors: $n_i = N_i = 5$.  The result, by Eqs.\ (\ref{eq:enf}) and (\ref{eq:ene}), is that every NC has the same two energy levels, $\varepsilon^{(f)} = E_Q(5) - e^2/\kappa D$ and $\varepsilon^{(e)} = E_Q(5) + e^2/\kappa D$, and the system's Fermi level $\mu = E_Q(5)$.  Equivalently, one can say that the single-particle DOGS for this hypothetical system corresponds to two $\delta$-function peaks at $\varepsilon = \pm e^2/\kappa D$.  

As explained in the introduction, conduction in this uniformly neutral system requires the excitation of a positive/negative NC pair.  Specifically, such an excitation produces one positive NC containing $4$ electrons and one negative NC containing $6$, and as such it has an excitation energy equal to the sum of the two Coulomb self-energies.  Equivalently, one can say that conduction requires the production of a hole in the filled $\delta$-function DOGS peak at $\varepsilon = -e^2/\kappa D$ and an electron in the empty DOGS peak at $\varepsilon = e^2/\kappa D$, and so the conduction has an activation energy $\varepsilon_A = \varepsilon_c = e^2/\kappa D$.  Thus, this hypothetical system without disorder has activated conduction: $\rho = \rho_0 \exp[\varepsilon_A/k_BT]$.

On the other hand, once the randomness in donor number is taken into account, one can no longer say in general that the ground state arrangement of electrons is uniformly neutral, $n_i = N_i$.  Indeed, when $N_i$ can take a wide range of values, then those NCs with very large $N$ may become ionized so that their electrons can occupy lower-energy shells on other NCs with small $N$.  In this way, the presence of a discrete quantum energy spectrum instigates the production of positively- and negatively-charged NCs.  It is this spontaneous charging that allows for VRH, as we will show below.

Still, it is straightforward to see that the system remains nearly uniformly electroneutral in the ground state under either of two conditions: (i) very small quantum energy gap, $\Delta \ll 1$, or (ii) very small doping level, $\nu \ll 1$.  In the former case, the difference between quantum energy levels becomes negligibly small compared to the energy required to produce charging of NCs.  Thus, the NCs remain neutral and the conduction is activated, as explained above.  In the limit of very small doping, $\nu \ll 1$, the system also remains nearly uniformly neutral due to an extreme scarcity of donors with $N_i > 2$.  Indeed, by Eq.\ (\ref{eq:Poisson}), at small $\nu$ the fraction of donors with $N_i > 2$ is $\simeq \nu^3/6$.  Thus, neutrality of the system can be maintained without requiring any significant number of electrons to occupy the 1P shell, and there is essentially no charging of NCs.  Therefore in the limit of very small $\nu$ the conduction is also activated.

In situations where either $\Delta$ or $\nu$ is not small, one can expect spontaneous charging of NCs in the ground state, and it is not trivial to predict the DOGS or the temperature dependence of the resistivity.  We explore these situations using our simulation method, outlined in Sec.\ \ref{sec:computer}.  Before proceeding to present results for a wide range of $\nu$ and $\Delta$, we first focus on the illustrative cases of $\nu = 5$ and $\nu = 2$, taking for the quantum energy gap $\Delta = 5$.

At $\nu = 5$, the Fermi level resides in the middle of the 1P shell.  Thus, since the gap between quantum energy levels is relatively large, in the ground state essentially all NCs satisfy $2 \leq n_i \leq 8$.  By Eq.\ (\ref{eq:Poisson}), however, roughly 11\% of NCs have a donor number satisfying $N_i < 2$ or $N_i > 8$.  Such NCs become charged in the ground state, driven by the large gaps in the quantum energy spectrum that induce electrons to leave the 1D shell and to fill the 1S shell.  Thus, the ground state configuration of the system consists of randomly-distributed fixed charges, which correspond to those NCs with $N_i < 2$ (which become negatively-charged) or $N_i > 8$ (positively-charged), and mobile electrons and holes in the partially-filled 1P shell.  The mobile electrons and holes arrange themselves in such a way that the ES criteria of Eqs.\ (\ref{eq:EScrit1}) and (\ref{eq:EScrit2}) are satisfied.  It is these criteria that give rise to the vanishing DOGS near the Fermi level \cite{Efros1975cga, Efros1984epo}.

\begin{figure}[htb!]
\centering
\includegraphics[width=0.45 \textwidth]{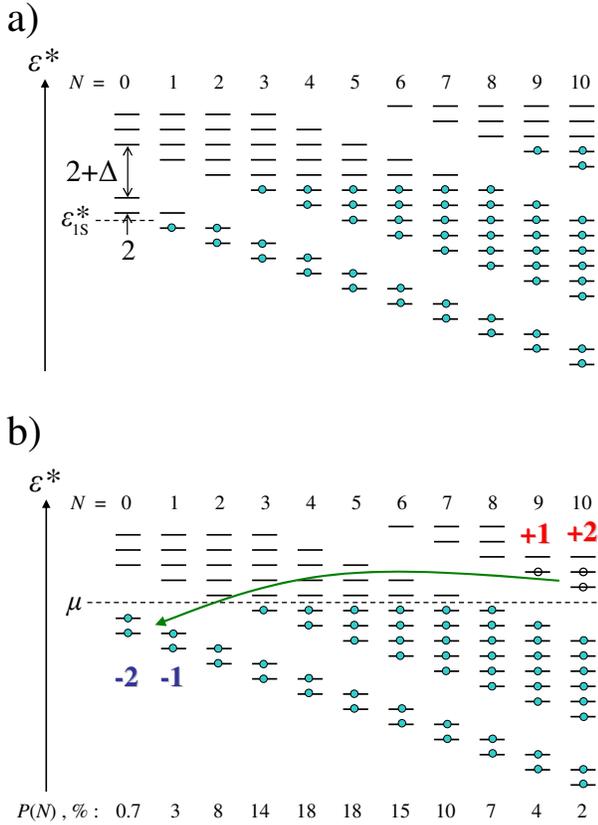}
\caption{(Color online) Schematic depiction of the charging process in a system with NCs with varying donor number $N$.  a) The single-electron energy levels (horizontal line segments) are shown for isolated NCs.  The Coulomb self-energy of charged NCs produces a spectrum where different charge states have a separation $2 e^2/\kappa D$.  The quantum confinement energy provides a gap between subsequent shells, e.g. 1S and 1P states or 1P and 1D states.  In the neutral state, a NC with $N$ donors has $N$ filled energy levels (indicated by filled blue dots).  $\varepsilon^{*}_\text{1S}$ indicates the quantum kinetic energy of the 1S shell, $\varepsilon^{*}_\text{1S} = E_Q(1)/(e^2/\kappa D)$. b) A depiction of the charging process at $\nu = 5$.  Electrons in the 1D shell of NCs with $N > 8$ abandon these NCs and instead fill empty energy levels in the 1S shell of NCs with $N < 2$.  In this way NCs with $N > 8$ become positively charged and NCs with $N < 2$ become negatively charged.  The resulting Fermi level $\mu$ is shown by the dashed line.  For NCs with $N = 5$, it resides in the center of the 1P shell.  The relative abundance of different donor numbers at $\nu = 5$ is shown at the bottom of the figure as a percentage.} \label{fig:energylevels}
\end{figure}

This process of charging of NCs is illustrated schematically in Fig.\ \ref{fig:energylevels}, which shows the energy levels of isolated NCs with donor numbers $0 \leq N \leq 10$.  In the neutral state, a NC with $N$ donors has $N$ filled electron energy levels (Fig.\ \ref{fig:energylevels}a).  When the system contains a mixture of NCs with different $N$, however, electrons abandon high energy levels in NCs with large $N$ and fill empty states in NCs with small $N$.  This process is shown for the case $\nu = 5$ in Fig. \ref{fig:energylevels}b.  The resulting charged NCs produce a random Coulomb potential throughout the system that smears the single electron energy levels and produces a finite density of states near the Fermi level.

\begin{figure}[htb!]
\centering
\includegraphics[width=0.5 \textwidth]{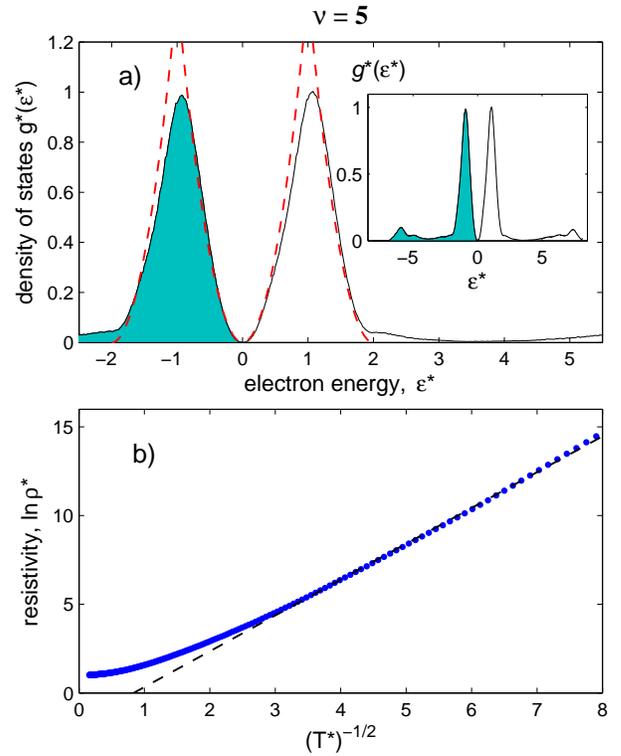}
\caption{(Color online) Density of ground states and resistivity at $\nu = 5$ and $\Delta = 5$, as measured by computer simulation.  a) Density of states as a function of electron energy.  Filled electron states are shaded.  Dashed red lines show, schematically, the quadratic Coulomb gap near the Fermi level, $\varepsilon^* = 0$, and the ``reflected Coulomb gaps" at $\varepsilon^* = \pm 2$.  Note that that the total shaded and unshaded areas under the $g^*(\varepsilon^*)$ curve are both normalized to unity, since each NC has one electron and one hole excitation.  The inset shows the DOGS over a wider energy range, with small, distant peaks indicating rare NCs whose highest filled electron state is in the 1S shell or whose first empty state is in the 1D shell.   b)  The dimensionless logarithm of the resistance, $\ln \rho^*$, as a function of $(T^*)^{-1/2}$, which illustrates the existence of ES resistivity at small temperature.} \label{fig:nu5}
\end{figure}

The DOGS for $\nu = 5$ and $\Delta = 5$, as calculated by our numerical simulation, is plotted in Fig.\ \ref{fig:nu5}a.  One can see the quadratic Coulomb gap near the Fermi level, as proscribed by the ES theory.  As compared to the conventional Coulomb gap problem in lightly-doped semiconductors \cite{Efros1984epo}, this Coulomb gap is remarkably well preserved, with the DOGS remaining quadratic until $\varepsilon^* \approx 1$.  This strong Coulomb gap suggests that the resistivity should follow the ES law for all temperatures $T^* \ll 1$.  Specifically, at these small temperatures the resistivity is described by
\be 
\rho(T) = \rho_0 \exp\left[ \left(\frac{T_{ES}}{T} \right)^{1/2} \right],
\label{eq:ES}
\ee
where 
\be 
T_{ES} = \frac{C e^2}{k_B \kappa \xi}
\label{eq:TES}
\ee 
and $C$ is a numerical coefficient of order unity.

This behavior can indeed be seen in Fig.\ \ref{fig:nu5}b, where $\ln \rho^*$ is plotted as a function of $(T^*)^{-1/2}$.  The linear relationship at large $(T^*)^{-1/2}$ suggests that, as expected, the resistance follows the ES law at small temperatures.  We find that the numerical coefficient $C \approx 8.1$, as compared to the typical value $C \approx 2.8$ in lightly-doped bulk semiconductors \cite{Efros1984epo}.  At larger temperatures $T^* > 1$ [or $(T^*)^{-1/2} < 1$], the resistivity saturates at $\ln \rho^* = 1$.  At such large temperatures the factor $\varepsilon_{ij}^*/T^*$ in Eq.\ (\ref{eq:lnrho}) typically becomes smaller than unity, which indicates that electrons tunnel relatively easily between nearest neighbors, and VRH is abandoned in favor of nearest-neighbor hopping.  At these large temperatures the resistivity can be expected to have only a relatively weak power-law dependence on temperature, which is beyond the accuracy of our numerical calculations.

In addition to the parabolic Coulomb gap near the Fermi level, another salient feature of the DOGS in Fig.\ \ref{fig:nu5}a is that it has strong maxima at $\varepsilon^* = \pm 1$ and collapses nearly to zero at $\varepsilon^* = \pm 2$, as if there were additional Coulomb gaps that constrain the density of states around $\varepsilon^* = \pm 2$.  These ``reflected Coulomb gaps" are in fact the product of an approximate symmetry in the system, which can be seen by examining Eqs.\ (\ref{eq:enf}) and (\ref{eq:ene}).  At $\nu = 5$, the great majority of NCs have $2 < n_i < 8$.  For such NCs, $E_Q(n_i) = E_Q(n_i + 1)$; both the highest filled and lowest empty electron states are in the 1P shell.  In this case, one can subtract Eqs.\ (\ref{eq:enf}) and (\ref{eq:ene}) to show that $\varepsilon_i^{*(e)} = \varepsilon_i^{*(f)} + 2$.  Thus, the great majority of NCs contribute to the density of states two energy levels -- one filled, one empty -- separated by $2 e^2/\kappa D$.  This creates an approximate discrete translational symmetry in the density of states, so that $g^*(\varepsilon^*) \approx g^*(\varepsilon^* - 2)$ for $0 < \varepsilon^* < 2$.  As a consequence, the Coulomb gap at the Fermi level implies the existence of reflected Coulomb gaps at $\varepsilon^* = \pm 2$.  In other words, one can say that because of the discrete charging spectrum of NCs the conventional quadratic bound on the DOGS near the Fermi level also produces (approximate) quadratic bounds on the DOGS near $\varepsilon^* = \pm 2$.  The contribution of rare NCs with $n_i = 2$ or $n_i = 8$ to the DOGS can be seen in the small peaks at $\varepsilon^* = -6$ and $\varepsilon^* = 7$, as shown in the inset of Fig.\ \ref{fig:nu5}a.

The presence of reflected Coulomb gaps is not unique to the doping level $\nu = 5$.  Indeed, for all $\nu$ that are sufficiently removed from the quantum energy gaps at $\nu = 2$, $\nu = 8$, etc., the relation $\varepsilon_i^{*(e)} = \varepsilon_i^{*(f)} + 2$ is valid for most NCs in the system and the resulting DOGS is essentially identical to that of Fig.\ \ref{fig:nu5}a.  Consequently, the resistivity plot shown in Fig.\ \ref{fig:nu5}b accurately describes the resistivity at most values of $\nu > 1$.  The reflected Coulomb gaps in Fig.\ \ref{fig:nu5}a appear even more dramatically for large NCs with external impurity charges, as will be shown in Sec.\ \ref{sec:largeoxide}.

\begin{figure}[htb!]
\centering
\includegraphics[width=0.4 \textwidth]{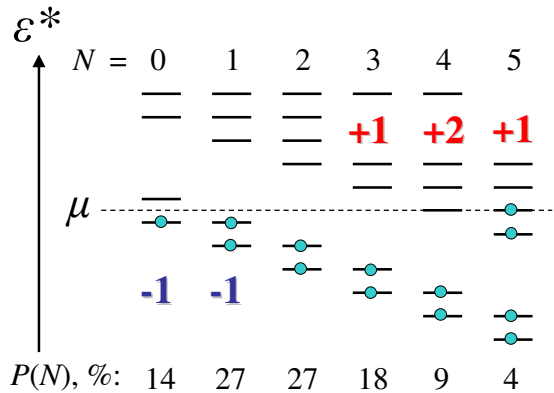}
\caption{(Color online) Schematic depiction of the filled and empty energy levels at $\nu = 2$.  Energy levels are shown for NCs in the absence of any Coulomb potential, similar to Fig.\ \ref{fig:energylevels}.  At $\nu = 2$, some electrons leave the 1P shell of NCs with $N > 2$ and fill empty states in the 1S shell of NCs with $N < 2$.  The resulting Fermi level $\mu$ is aligned with the first(second) energy level of the 1P shell in NCs with $N = 4$(5), which is partially filled.} \label{fig:energylevelsnu2}
\end{figure}

On the other hand, one could expect qualitatively different behavior at $\nu = 2$, where there are precisely enough electrons to fill the 1S shell of every NC, and the Fermi level sits in between the 1S and 1P shells.  In this case there is no ``discrete translational symmetry" in the density of states, since the empty and filled energy levels for most NCs, $\varepsilon_i^{(e)}$ and $\varepsilon_i^{(f)}$, sit on opposite sides of the quantum energy gap, as shown schematically in Fig.\ \ref{fig:energylevelsnu2}.  This produces a DOGS that is qualitatively different from what is shown in Fig.\ \ref{fig:nu5}a.  One could therefore expect that the dependence of the resistivity on temperature is also qualitatively different.  Such thinking is supported by a recent experiment on electrochemically gated NCs \cite{Liu2010mae}, which reported that when $\nu$ is very close to $2$ there appears an appreciable temperature window over which the resistivity follows the Mott law.  Given these differences, it is worth giving some special consideration to the case $\nu = 2$.

\begin{figure}[htb!]
\centering
\includegraphics[width=0.5 \textwidth]{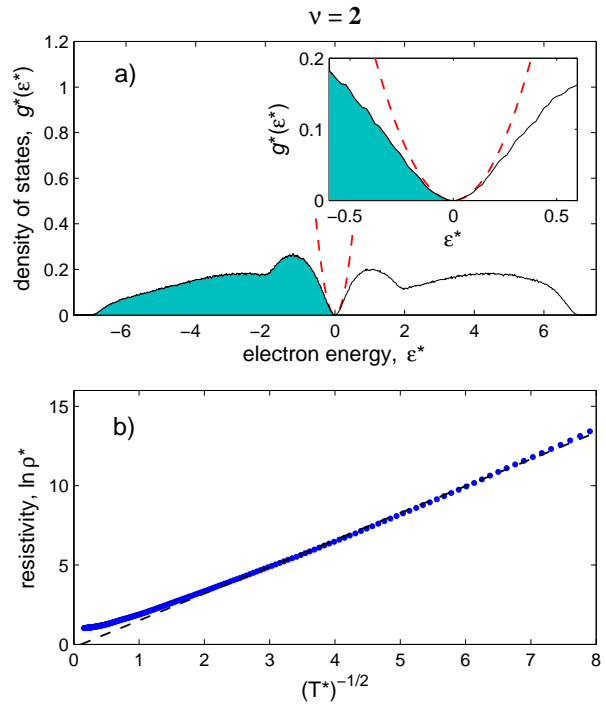}
\caption{(Color online) Density of states and resistivity at $\nu = 2$ and $\Delta = 5$, as measured by computer simulation.  a) DOGS as a function of electron energy.  Filled electron states are shaded.  The dashed red curve is the same parabolic curve shown in Fig.\ \ref{fig:nu5}a.  The inset shows the DOGS very close to the Fermi level. b)  The dimensionless logarithm of the resistance, $\ln \rho^*$, as a function of $(T^*)^{-1/2}$, which shows ES resistivity at $T^* \ll 1$.} \label{fig:nu2}
\end{figure}

The DOGS for $\nu = 2$ is shown in Fig.\ \ref{fig:nu2}a.  
Unlike at $\nu = 5$, where the DOGS collapses at $\varepsilon^* = \pm 2$, the DOGS at $\nu = 2$ is much broader, with a width $\Delta + 2$.  This broad DOGS can be seen as a consequence of the large gap between 1S and 1P energy shells, which implies that the energy of electron or hole excitations, $\varepsilon^{(f)}_i$ and $\varepsilon^{(e)}_i$, can take a wide range of values, depending on the donor number $N_i$.  Alternatively, one can say that since both 1S and 1P electron states contribute to the DOGS near the Fermi level, the density of states has a characteristic width similar to that of the gap $\Delta$.

As at $\nu = 5$, the DOGS vanishes at the Fermi level (see the inset of Fig.\ \ref{fig:nu2}a), but in this case it can only be described as parabolic over the fairly narrow range of energies $|\varepsilon^*| < 0.2$.  In the intermediate range of energies $0.2 < |\varepsilon^*| < 1$, the DOGS grows roughly linearly with energy.  At larger energies $1 < \varepsilon^* < \Delta$ the DOGS becomes roughly constant.

In spite of this relatively complicated DOGS, Fig.\ \ref{fig:nu2}b shows that the resistivity is in excellent agreement with the ES law, with a coefficient $C \approx 5.7$ [see Eq.\ (\ref{eq:TES})], at all but very large temperatures.  This is somewhat surprising, since it suggests that the system exhibits ES resistivity even when the temperature is large enough that the band of energies over which VRH occurs is much larger than the width of the parabolic Coulomb gap.  This behavior would be impossible if states were randomly distributed in space.  Our observation of ES resistivity suggests that at $\nu = 2$ spatial correlations emerge which somehow preserve ES resistivity even in the absence of a parabolic DOGS.  

To illustrate how this might be possible, let us first recall that in a disordered two-dimensional (2D) system, the DOGS is linear in energy near the Fermi level rather than parabolic, but the ES law of VRH is still obeyed \cite{Efros1975cga}.  One can now imagine a 3D system in which sites with energies close to the Fermi level are arranged in a 2D fractal subspace embedded in the system volume.  In such a system, one would still have a linear DOGS near the Fermi level accompanied by ES resistivity, even though the system as a whole is three-dimensional.  Using this reasoning, one can speculate that the results shown in Fig.\ \ref{fig:nu2} are indicative of such a fractal arrangement of sites near the Fermi level, driven in some way by the long-ranged Coulomb potential.  More broadly, these results hint at the idea that in a disordered system of localized states dominated by Coulomb interactions, one should be able to derive the ES law without explicit reference to the DOGS or the system's dimensionality.  Such an argument was in fact first put forward by Larkin and Khmelnitskii \cite{Larkin1982aci}.  Our system at $\nu = 2$ may be a good application of this argument.  It remains unclear, however, in which situations this argument is applicable \textit{a priori}.  This general question and its application to the case $\nu = 2$ will be the subject of a future publication.

By conventional thinking, the relatively constant DOGS at $|\varepsilon^*| > 1$ would seem to suggest a regime of temperature in which the resistivity follows the Mott law, which describes VRH in the presence of a constant DOGS.  However, unlike the experiments of Ref.\ \cite{Liu2010mae}, we see no noticeable region of Mott VRH.  The Mott resistivity observed in Ref.\ \cite{Liu2010mae} at $\nu = 2$ is likely the result of some additional disorder that is outside the model considered in this section, and is discussed further in Sec.\ \ref{sec:IL}.

Having considered the specific cases of $\nu = 5$ and $\nu = 2$, we now turn our attention to a general description of VRH at different values of $\nu$ and $\Delta$.  In order to identify more precisely which conditions produce VRH, we used our simulation to measure the resistivity as a function of $T^*$, $\nu$, and $\Delta$ over the range $0.01 \leq T^* \leq 10$, $0.2 \leq \nu \leq 2$, and $0.5 \leq \Delta \leq 5$.  For each case we measured the exponent $\gamma$ of the temperature dependence of resistivity by calculating the ``reduced activation energy" $w(T^*) = - d(\ln \rho^*)/d(\ln T^*) \propto T^{-\gamma}$ \cite{Zabrodskii1977}.  The exponent $\gamma$ was identified by making a power law best fit to $w(T^*)$.
Those values of $T^*$, $\nu$, and $\Delta$ that produce $\gamma = 0.5 \pm 0.1$ were identified with ES resistivity; domains where $\gamma > 0.6$ were identified with activated resistivity.  As discussed above, no significant regimes were identified that showed Mott behavior.  We use this data to construct an approximate phase diagram in the space of $T^*$, $\nu$, and $\Delta$ that identifies which behavior can be expected.

Our result is plotted in Fig.\ \ref{fig:phasediagram} for $T^* \ll 1$.  Generally speaking, the results indicate that for $\nu > 0.6$ and $\Delta > 0.5$ one can expect ES resistivity, while for other conditions the resistivity is activated.  These conditions are equivalent to the conditions (i) and (ii) that were announced in the introduction.  Dashed horizontal lines indicate, as an example, the values of $\Delta$ corresponding to CdSe NCs with $D = 6.2$ nm, as in Ref.\ \cite{Liu2010mae}, and Si NCs with $D = 5$ nm, as in Ref. \cite{Jurbergs2006snw}.  Both of these dashed lines assume that $\knc/\ki = 5$.  At temperatures $T^* > 1$ VRH is gradually replaced by nearest-neighbor hopping.  The condition $T^* < 1$ is equivalent to the condition (iii) from the introduction.

As mentioned above, the model considered in this section does not account explicitly for any sources of disorder other than fluctuations in donor number.  For example, in real NC arrays the diameter $D$ varies from one NC to another, which introduces variations in the quantum spectrum between NCs [see Eq.\ (\ref{eq:EQ})].  Nonetheless, the presence of these size fluctuations in addition to fluctuations in donor number does not destroy ES VRH, since the Coulomb gap near the Fermi level is a universal result of the ES stability criteria [Eqs.\ (\ref{eq:EScrit1}) and (\ref{eq:EScrit2})] and is independent of the source of disorder in the system.  Whether size fluctuations or other sources of disorder enhance the role of VRH or significantly affect the magnitude of the resistivity remains yet to be studied.  Generally speaking, however, one can expect that the phase diagram of Fig.\ \ref{fig:phasediagram} is accurate whenever the typical magnitude of size fluctuations $\delta D$ satisfies $(\delta D)/D \ll 1/\Delta$.  We further expect that even larger size fluctuations do not greatly affect VRH in regimes where the ES law applies, since in such cases the DOGS is already saturated by the disorder in donor number.  In regimes where the resistivity is activated, the presence of a large additional disorder should generally promote the existence of VRH, which decreases the resistivity at small $T^*$.

\section{VRH in arrays of large NCs with external impurity charges} \label{sec:largeoxide}

In the previous sections we showed that the spontaneous charging that leads to VRH is driven by the relatively large gaps between degenerate shells of the electron quantum energy spectrum.  In large NCs, the gap $\Delta$ becomes small and this charging disappears, which leads to activated resistivity (see Fig.\ \ref{fig:phasediagram}).  A similar effect can be expected when the NC shape is not symmetric.  In this case, the electron energy levels are not degenerate, so that the bunch of energy levels corresponding to a particular shell in a spherical NC is dispersed, and as a consequence the gaps between subsequent energy levels are reduced.  Thus, large or highly asymmetric NCs tend to remain neutral in the ground state and exhibit activated transport even at $\nu > 1$.

In this section, however, we show that if donor impurities are located \emph{outside} of NCs, ES VRH can still be observed.  The presence of ES VRH in large NCs can be understood using an argument that was first put forward by Ref.\ \onlinecite{Zhang2004dos} in the context of granular metallic films.  The argument from Ref.\ \onlinecite{Zhang2004dos} is briefly repeated here.  

\begin{figure}[htb!]
\centering
\includegraphics[width=0.4 \textwidth]{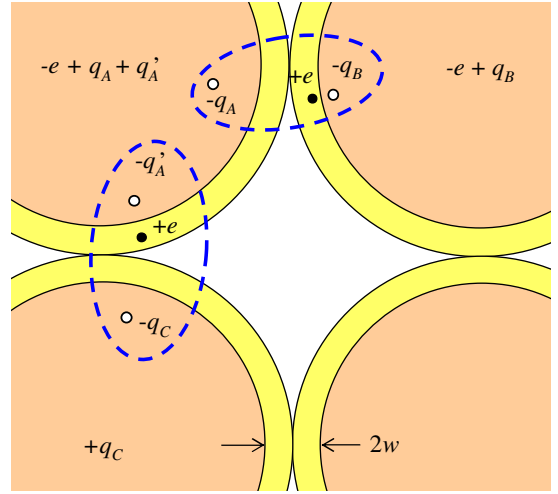}
\caption{(Color online) A schematic depiction of the fractionalization of the charge of a donor impurity between large NCs (see also Fig.\ 11 of Ref.\ \cite{Zhang2004dos}).  Semiconductor NCs (tan/gray circles) have a thin coating of insulator (yellow/light gray), with embedded donor impurities (small black circles).  Each positive donor induces negative image charges (small white circles) that neutralize it, while equal and opposite positive images are conveyed to the center of the NC.  Those donors that are located close to the point of contact between two NCs create non-integer image charges in the two surrounding NC surfaces.  In this way NCs are given a ``fractional donor charge" $Q_i$.} \label{fig:metalschematic}
\end{figure}

Consider an array of spherical semiconductor NCs with large internal dielectric constant $\knc$, each of which is coated with a thin layer of width $w$ of insulator, such as the semiconductor's own oxide or the ligands shown in Fig.\ \ref{fig:schematic}.  Suppose further that donor impurity charges $+e$ are embedded in this insulator, as shown schematically in Fig.\ \ref{fig:metalschematic}, with some overall concentration $\Nox$.  If some particular donor resides at a location within the insulator shell that is well-separated from the points of contact between neighboring NCs, then this donor simply donates its electron to the NC on which it resides.  The resulting positive impurity charge induces a negative image charge on the NC surface because of the dielectric response, and together the donor and its image charge make a compact neutral pair.  In this way, donors that are not near the point of contact between two NCs produce a negligible Coulomb potential that plays no role in charging the system.

On the other hand, when a donor is located close to the point of contact between two NCs, labeled A and B, it induces negative image charges $-q_A$ and $-q_B$ in the surfaces of NCs A and B, respectively.  In order to maintain overall neutrality of the NCs, an equal and opposite image charge appears at the center of each NC: $+q_A$ and $+q_B$.  (These ``image charges at the center" represent a uniform electronic charge at the NC surface.)  The values of $q_A$ and $q_B$ are such that together the image charges $-q_A$ and $-q_B$ neutralize the donor charge: $q_A + q_B = e$.  Their respective magnitudes are determined by the distance between the impurity and each NC surface.  For example, if the impurity sits exactly along the line connecting the centers of NCs A and B and if the gap $2w$ between NCs satisfies $w \ll D$, then the NCs can be approximated as infinite planar metallic surfaces and $q_A x_B = q_B x_A$, where $x_A$ and $x_B$ are the distances between the impurity and the surface of NCs A and B, respectively.  More generally, when $w \ll D$, the image charges $-q_A$ and $-q_B$ sit very close to the donor impurity and essentially neutralize it, so that the screened impurity does not directly contribute to any Coulomb potential at length scales of $D$ or larger.  Instead, the net effect of the image charges is to ``fractionalize" the donor impurity charge, such that $+q_A$ is relayed to the center of NC A and $+q_B$ is relayed to the center of B.  This process is depicted in Fig.\ \ref{fig:metalschematic}.  

In this way, each NC gets a number of random, positive fractional donor charges created by those donors located near the contact points between NCs.  We denote the sum of all fractional charges at NC $i$ by $Q_i$.
The proportion of all donor charges that sit at these contact points is $\sim w/D$, so that of the total number $\sim \Nox w D^2$ of donor impurities in the insulator shell covering a given NC, only $\sim \Nox w^2 D$ of these become fractionalized.  When the average total number of fractionalized charges per NC $\Nox w^2 D \gg 1$, the central limit theorem guarantees that the distribution of the random variable $Q_i/e$ can be approximated as a Gaussian with mean $\sim \Nox w^2 D$ and root mean square fluctuation $\sim(\Nox w^2 D)^{1/2}$.  

Donor electrons respond to the potential created by fractional charges by arranging themselves on NCs in integer number and in such a way that the total electrostatic energy of the system is minimized.  In other words, in the ground state the set of electron occupation numbers $\{n_i\}$ is that which minimizes the Hamiltonian
\be
H = \sum_i \frac{(Q_i - en_i)^2 }{\kappa D} 
+ \sum_{\langle i,j \rangle } \frac{(Q_i - en_i)(Q_j - en_j)}{\kappa r_{ij}}.
\label{eq:Hmetal}
\ee 
Unlike in the Hamiltonian of Eq.\ (\ref{eq:H}) of Sec.\ \ref{sec:model}, here the quantum energy gaps are negligibly small and disorder is provided by the fractional charges $\{Q_i\}$.  

Eq.\ (\ref{eq:Hmetal}) implies that the corresponding electron energy levels at NC $i$ are given by
\be
\varepsilon_i^{(f)} = \frac{(Q_i - en_i)^2 - (Q_i - en_i + e)^2}{\kappa D} - e\sum_{j \neq i} \frac{Q_j - en_j}{\kappa r_{ij}} 
\label{eq:enfmetal}
\ee
and
\be
\varepsilon_i^{(e)} = \frac{(Q_i - en_i - e)^2 - (Q_i - en_i)^2}{\kappa D} - e\sum_{j \neq i} \frac{Q_j - en_j}{\kappa r_{ij}},
\label{eq:enemetal}
\ee
for the highest filled and lowest empty states, respectively.  By subtracting Eqs.\ (\ref{eq:enfmetal}) and (\ref{eq:enemetal}) it can be seen that $\varepsilon^{(e)} = \varepsilon^{(f)} + 2e^2/\kappa D$ for all NCs.  This has important implications for the DOGS, as will be shown below.

As explained above, the values of the fractional charges $Q_i$ can be assigned using a Gaussian distribution with mean $\bar{Q} \sim e\Nox w^2 D$ and standard deviation $\sigma_Q = (e\bar{Q})^{1/2}$.  One can notice, however, that in the Hamiltonian of Eq.\ (\ref{eq:Hmetal}) the variables $Q_i$ and $n_i$ appear only in the combination $Q_i - e n_i$, which by electroneutrality of the system must satisfy $\langle Q_i - e n_i \rangle = 0$.  Thus, when calculating the DOGS and resistivity, one can adopt a somewhat simpler model where $Q_i$ is chosen from a distribution with mean zero and $n_i$ is allowed to take any integer value (positive or negative).

In fact, an even further simplification of the model is available when the standard deviation $\sigma_Q/e \gg 1$.  Namely, when $Q_i$ can take such a wide range of values, one can approximately replace the broad distribution for $Q_i/e$ with a uniform distribution $Q_i/e \in [-1/2, +1/2]$.  The validity of this approximation can be understood by considering that each NC minimizes its Coulomb self-energy by minimizing the magnitude of its net charge, $|Q_i - en_i|$.  Since $n_i$ can take any integer value, it is generally true that in the ground state $-e/2 \leq Q_i - e n_i \leq e/2$.  This random spatial arrangement of net charges produces a fluctuating Coulomb potential that leads to ES VRH
\footnote{It can be noted that randomly choosing the donor charge $Q_i$ from the interval $[-e/2, +e/2]$ does not result in a system whose overall donor charge is neutral for a given realization.  One could alternatively choose the set $\{Q_i\}$ in such a way that the total donor charge is zero, i.e. $\sum_i Q_i = 0$.  In our simulation we find that introducing this constraint provides no noticeable effect on either the DOGS or the resistivity.}.
All results below correspond to this choice of a uniform distribution for $Q_i$.

It is worth noting that while so far we have focused on positive donor impurities, acceptors embedded in the insulating layer can play the same role.  Random charging of NCs can also result from the simultaneous presence of both donors and acceptors.  For the sake of argument, however, we focus our discussion around positive donor charges.

\begin{figure}[htb!]
\centering
\includegraphics[width=0.5 \textwidth]{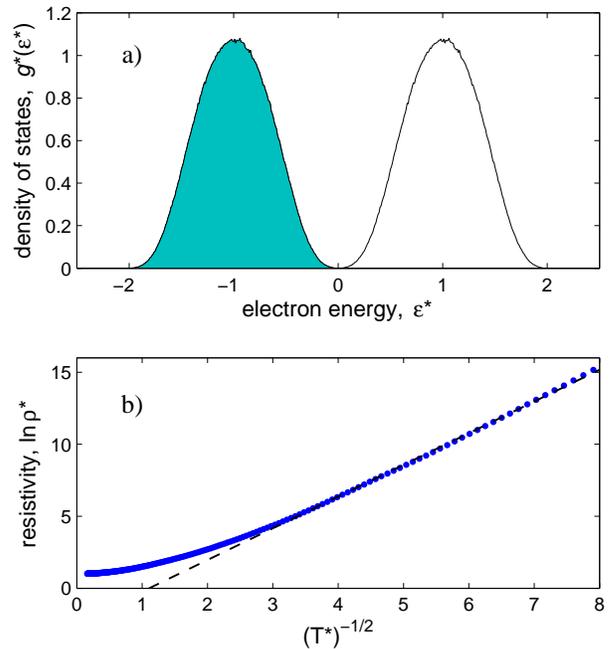}
\caption{(Color online) Electron DOGS and resistivity for an array of large semiconductor NCs with fractional donor charges corresponding to $Q_i/e \in [-1/2, +1/2]$.  a) The DOGS, which vanishes at $\varepsilon^* = 0$ and at $\varepsilon^* = \pm 2$ because of the ES stability criteria and the relation between filled and empty energy states at each NC: $\varepsilon_i^{*(e)} = \varepsilon_i^{*(f)} + 2$.  b) Resistivity versus $(T^*)^{-1/2}$, which shows ES behavior at temperature $T^* \ll 1$.} \label{fig:metal}
\end{figure}

Results for the DOGS and resistivity are given in Fig.\ \ref{fig:metal}, as calculated using the simulation method described in Sec.\ \ref{sec:computer} and the definition of single-particle energies given in Eqs.\ (\ref{eq:enfmetal}) and (\ref{eq:enemetal}).  One can note that the DOGS shown in Fig.\ \ref{fig:metal}a vanishes at the Fermi level $\varepsilon^* = 0$, as required by the ES criteria [Eqs.\ (\ref{eq:EScrit1}) and (\ref{eq:EScrit2})], as well as at $\varepsilon^* = \pm 2$.  Because of the lack of a quantum energy term in the Hamiltonian, the DOGS also has a perfect discrete translational symmetry: $g^*(\varepsilon^*) = g^*(\varepsilon - 2)$ for $\varepsilon^* > 0$.  
The strong ``reflected Coulomb gaps" in Fig.\ \ref{fig:metal} are a result of the relation $\varepsilon_i^{*(e)} = \varepsilon_i^{*(f)} + 2$. 

The resistivity for this system is plotted in Fig.\ \ref{fig:metal}b as a function of $(T^*)^{-1/2}$.  As expected, the resistivity follows the ES law at $T^* \ll 1$ due to the strongly-preserved quadratic Coulomb gap near the Fermi level.  The dashed line in Fig.\ \ref{fig:metal}b corresponds to the ES law with a coefficient $C \approx 9.6$ [see Eq.\ (\ref{eq:TES})].

We have also verified that our results for the DOGS and resistivity are practically identical if $Q_i/e$ is chosen not from a uniform distribution $[-1/2, +1/2]$ but from a Gaussian distribution with three-times larger variance.

In the opposite limit, where fractionalized donor charges are very rare, $\Nox w^2 D \ll 1$, each NC remains essentially neutral, and there is no random Coulomb potential.  This uniformity leads to activated nearest-neighbor hopping, since without disorder long-range electron hops cannot reduce the energy required for the hop
\footnote{Here, following the main ideas of Ref.\ \onlinecite{Zhang2004dos}, we disagree with the estimates in the last paragraph of Ref.\ \onlinecite{Zhang2004dos}'s Sec.\ V.  Those estimates are appropriate for insulator-coated spheres with distance $d \gtrsim D/2$ from each other, but not for much closer touching spheres with $d = 2w$.  In the latter case, the regions of close contact between neighboring spheres play the special role of creating fractionalized donor charges, as discussed in Sec.\ \ref{sec:largeoxide}.}.

Finally, it can be noted that in our discussion above we have ignored the possible presence of deep electronic states at the NC surface.  Such trap states can play the role of compensating impurity centers, which remove a percentage of electrons from the conduction band.  In this case, only the uncompensated donor electrons contribute to conduction, and the value of $\nu$ is effectively renormalized downward.  Repulsion between uncompensated donor electrons and electrons in trap states may also produce a small shift in electron energies, and is outside the scope of our treatment here.

\section{Gating of a NC array by an ionic liquid} \label{sec:IL}

In Secs.\ \ref{sec:intro} -- \ref{sec:largeoxide} we discussed systems of NCs doped by random impurities, and we explored the dependence of the resistivity on the doping level.  In such systems, the doping level is established during the fabrication of NCs.  In many cases, however, it is desirable to have a doping level that can be continuously tuned, so that the resistivity of a single device can be set to a wide range of values.  For this purpose, electrochemically gated arrays of semiconductor NCs are actively being studied \cite{Liu2010mae, Yu2003ncc}.  

\begin{figure}[htb!]
\centering
\includegraphics[width=0.4 \textwidth]{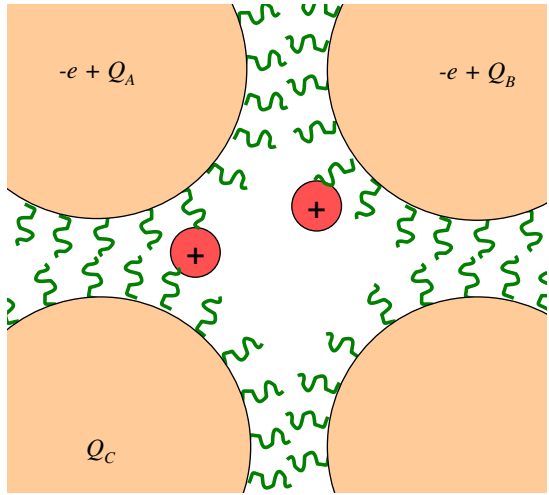}
\caption{(Color online) A schematic picture of an array of semiconductor NCs (large circles) gated by an ionic liquid.  Cations (small circles with $+$'s) are driven by a voltage source to intercalate between NCs.  Because of the large NC dielectric constant $\knc$, the net effect of positive ions is to provide a fractional donor charge $Q_i$ at a given NC $i$, similar to what is shown in Fig.\ \ref{fig:metalschematic}.  Neutralizing electrons occupy NCs in order to neutralize ionic charges.  Ligands separating NCs are shown as curvy lines.} \label{fig:ILschematic}
\end{figure}

In such systems, conduction electrons are introduced into the system via a voltage source, which drives electrons from a top gate to a bottom gate that is in electrical contact with the NC array.  Generally, in between the top gate and the NC array is a room temperature ionic liquid that provides large capacitance and therefore allows for a high density of electrons to be introduced to the NC array at a relatively small voltage \cite{Chen2011cci}.  The cations from this ionic liquid intercalate into the spaces between NCs, penetrating deep into the array through the percolating network of pores between NCs, and thus provide a neutralizing charge for the conduction electrons.  A schematic picture of this system is given in Fig.\ \ref{fig:ILschematic}.

The large internal dielectric constant of NCs and the relatively small diameter of cations suggests the presence of strong image charge forces that bind cations electrostatically to their image charges in the NC surface.  In this way, one can expect that cationic charges are located primarily on the surface of each NC.  If one assumes that the position of cations on the NC surfaces is random, then one again arrives at a model of fractionalized cation image charges, similar to what is suggested in Sec.\ \ref{sec:largeoxide}.

For this model one can use a Hamiltonian that includes both a prominent quantum kinetic energy spectrum, as in Sec.\ \ref{sec:model}, and a fluctuating, fractionalized donor charge, as in Sec.\ \ref{sec:largeoxide}:
\begin{eqnarray}
H & = & \sum_i \left[ \frac{(Q_i - en_i)^2 }{\kappa D} + \sum_{k = 0}^{n_i}E_Q(k) \right] \nonumber \\
& & + \sum_{\langle i,j \rangle } \frac{(Q_i - en_i)(Q_j - en_j)}{\kappa r_{ij}}
\label{eq:HIL}
\end{eqnarray}
Here, the fractional charge $Q_i/e$ can be chosen uniformly from the interval $[\nu - 1/2, \nu + 1/2]$.  

Using our computer simulation method, we have briefly investigated the DOGS and resistivity of the system described by this Hamiltonian at various values of $\nu \geq 1$.  We find that ES VRH appears at low temperature for all values of $\nu > 1$.  In fact, when $|\nu - 2| > 1$ and $|\nu - 8| > 1$,  the DOGS is exactly the same as in Fig.\ \ref{fig:metal}a, and the resistivity is also identical.

We note that the model defined by Eq.\ (\ref{eq:HIL}), where the fractional donor charge is completely random, is unlikely to be accurate when $\nu$ is at the boundary between two quantum energy shells.  At $\nu = 2$, for example, random fractional charges lead to a fluctuating Coulomb potential with characteristic amplitude much larger than $k_BT/e$ at room temperature.  However, such a large Coulomb potential induces cations, which are mobile during the gating process, to rearrange in order to screen the potential.  In this way the cation positions become correlated and the typical amplitude of the Coulomb potential is reduced to $k_BT/e$, which is not large enough to produce charging of NCs.  As a result, the typical amplitude of fluctuations in $Q_i$ is likely much smaller than $e$, so that one should not expect a finite DOGS near the Fermi level.  Rather, in the absence of any other disorder, the resistivity should be large and activated.

\begin{figure}[htb!]
\centering
\includegraphics[width=0.45 \textwidth]{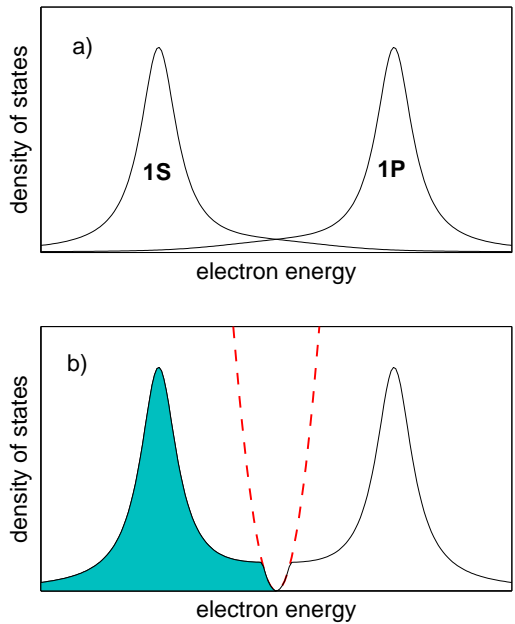}
\caption{(Color online) Schematic picture of the density of states at $\nu = 2$ in the presence of fluctuations in the NC diameter $D$.  a) If $D$ has some wide-tailed distribution, then the 1S and 1P energy levels are broadened and have a finite overlap.  b) Spatial correlations between rare 1S and 1P energy states near the Fermi level produce a Coulomb gap, so that ES resistivity is seen at very small temperatures and Mott resistivity is seen at larger temperatures.  } \label{fig:nu2schematic}
\end{figure}

Experiments with ionic liquid gating confirm that, as expected, the resistivity is much larger at $\nu = 2$ than at other filling factors \cite{Liu2010mae}.  However, the resistivity is generally shown to correspond to VRH rather than activated behavior, with ES resistivity seen at very small temperature and Mott resistivity at larger temperatures.  This VRH is likely the result of some other source of disorder, unrelated to the positions of cations, which produces finite DOGS near the Fermi level even at $\nu = 2$.  For example, if the NC diameters are not uniform, but are drawn from some distribution with finite width, then the energy levels corresponding to the 1S and 1P states are smeared.  If the distribution of NC diameters has wide tails, then the 1S and 1P energy levels can be smeared as far as the Fermi level, producing a finite DOGS near the Fermi level, as shown schematically in Fig.\ \ref{fig:nu2schematic}.

The overlap between some 1S and 1P energy levels produces rare NCs with $n = 3$ or $n = 1$ whose energy is very close to the Fermi level.  Such rare, mobile electrons are free to rearrange themselves in order to satisfy the ES stability criteria, and in doing so they produce a small Coulomb gap at the Fermi level (see Fig.\ \ref{fig:nu2schematic}).  As a result, the resistivity follows the ES law at very small $T$, and the Mott law at larger $T$, where the DOGS sampled by electron hops is essentially constant.  This is precisely what is seen in experiment \cite{Liu2010mae}.  

It is worth mentioning that ionic liquid gating of NC arrays allows one to measure the total electronic charge $Q$ as a function of applied gate voltage, or, in other words, the differential capacitance of the array $C = dQ/dV$.  In arrays of small spherical NCs, where the quantum gaps $\Delta$ dominate over Coulomb energies, most electrons enter the array when the voltage coincides with the energies of a quantum energy shell (1S or 1P, for example).  At such voltages the differential capacitance should have prominent peaks.  Between these voltages the capacitance should be small, reflecting the small electron DOGS.  We are not aware of any such experimental data \footnote{In our recent paper \cite{Chen2011cci}, we studied the hypothetical case where cations are large enough that only one cation can enter a pore in the NC array.  In this case, due to the Coulomb interaction, the cations form a crystal structure within the pores of the crystalline NC array.  This situation is different from the model considered in Sec.\ \ref{sec:largeoxide}, where ions are small and are introduced at relatively large temperature.  In Ref.\ \cite{Chen2011cci} we argued that in the former case the peak in capacitance corresponding to the 1S shell splits into two delta-function-like peaks, such that one electron enters every NC at two particular values of the voltage.}.

\section{Conclusions} \label{sec:conclusion}

In this paper we have used a simple theoretical model and a computer simulation to show how both activated transport and VRH arise in arrays of doped semiconductor NCs.  Our primary result is illustrated in the phase diagram of Fig.\ \ref{fig:phasediagram}: when the doping level $\nu$ and the quantum confinement energy $\Delta$ are sufficiently large, and when the temperature $T^*$ is sufficiently small, the resistivity of the array is characterized by ES VRH.  Such VRH is driven by the fluctuations in donor number from one NC to another, which lead to spontaneous charging of NCs as electrons depopulate higher quantum energy shells and fill lower ones.

We have also identified a striking feature of the DOGS in NC arrays: the presence of ``reflected Coulomb gaps" at electron energies $\pm 2 e^2/\kappa D$, which are a consequence of the ES stability criteria and the discrete charging spectrum of NCs (see Fig.\ \ref{fig:nu5}).  This feature is even more prominent in  large NCs with external impurity charges (Fig.\ \ref{fig:metal}).

The effect of additional disorder, such as fluctuations in NC size, remains yet to be explored quantitatively.  We conjecture, however, that for chemically doped NCs our results will be largely unaltered by the addition of such disorder.  For the case of NCs gated by ionic liquid, this external disorder seems crucial only for explaining the presence of Mott VRH at particular values of $\nu$ (see Fig.\ \ref{fig:nu2schematic}).

\begin{acknowledgments}

The authors would like to thank E. Aydil, Al. Efros, A. Efros, D. Frisbie, A. Frydman, Yu. Galperin, M. Goethe, P. Guyot-Sionnest, R. Holmes, J. Kakalios, A. Kamenev, U. Kortshagen, H. Liu, A. M\"{o}bius, A. Mkhoyan, M. Mueller, D. Norris, M. Palassini, M. Sung, and L. Wienkes for helpful discussions.
This work was supported primarily by the MRSEC Program of the National Science Foundation under Award Number DMR-0819885.  T. Chen was partially supported by the FTPI.

\end{acknowledgments}

\bibliography{random_doping-sc}

\begin{thebibliography}{29}%
\makeatletter
\providecommand \@ifxundefined [1]{%
 \@ifx{#1\undefined}
}%
\providecommand \@ifnum [1]{%
 \ifnum #1\expandafter \@firstoftwo
 \else \expandafter \@secondoftwo
 \fi
}%
\providecommand \@ifx [1]{%
 \ifx #1\expandafter \@firstoftwo
 \else \expandafter \@secondoftwo
 \fi
}%
\providecommand \natexlab [1]{#1}%
\providecommand \enquote  [1]{``#1''}%
\providecommand \bibnamefont  [1]{#1}%
\providecommand \bibfnamefont [1]{#1}%
\providecommand \citenamefont [1]{#1}%
\providecommand \href@noop [0]{\@secondoftwo}%
\providecommand \href [0]{\begingroup \@sanitize@url \@href}%
\providecommand \@href[1]{\@@startlink{#1}\@@href}%
\providecommand \@@href[1]{\endgroup#1\@@endlink}%
\providecommand \@sanitize@url [0]{\catcode `\\12\catcode `\$12\catcode
  `\&12\catcode `\#12\catcode `\^12\catcode `\_12\catcode `\%12\relax}%
\providecommand \@@startlink[1]{}%
\providecommand \@@endlink[0]{}%
\providecommand \url  [0]{\begingroup\@sanitize@url \@url }%
\providecommand \@url [1]{\endgroup\@href {#1}{\urlprefix }}%
\providecommand \urlprefix  [0]{URL }%
\providecommand \Eprint [0]{\href }%
\providecommand \doibase [0]{http://dx.doi.org/}%
\providecommand \selectlanguage [0]{\@gobble}%
\providecommand \bibinfo  [0]{\@secondoftwo}%
\providecommand \bibfield  [0]{\@secondoftwo}%
\providecommand \translation [1]{[#1]}%
\providecommand \BibitemOpen [0]{}%
\providecommand \bibitemStop [0]{}%
\providecommand \bibitemNoStop [0]{.\EOS\space}%
\providecommand \EOS [0]{\spacefactor3000\relax}%
\providecommand \BibitemShut  [1]{\csname bibitem#1\endcsname}%
\let\auto@bib@innerbib\@empty
\bibitem [{\citenamefont {Jurbergs}\ \emph {et~al.}(2006)\citenamefont
  {Jurbergs}, \citenamefont {Rogojina}, \citenamefont {Mangolini},\ and\
  \citenamefont {Kortshagen}}]{Jurbergs2006snw}%
  \BibitemOpen
  \bibfield  {author} {\bibinfo {author} {\bibfnamefont {D.}~\bibnamefont
  {Jurbergs}}, \bibinfo {author} {\bibfnamefont {E.}~\bibnamefont {Rogojina}},
  \bibinfo {author} {\bibfnamefont {L.}~\bibnamefont {Mangolini}}, \ and\
  \bibinfo {author} {\bibfnamefont {U.}~\bibnamefont {Kortshagen}},\ }\href
  {\doibase 10.1063/1.2210788} {\bibfield  {journal} {\bibinfo  {journal}
  {Applied Physics Letters}\ }\textbf {\bibinfo {volume} {88}},\ \bibinfo {eid}
  {233116} (\bibinfo {year} {2006})}\BibitemShut {NoStop}%
\bibitem [{\citenamefont {Rafiq}\ \emph {et~al.}(2006)\citenamefont {Rafiq},
  \citenamefont {Tsuchiya}, \citenamefont {Mizuta}, \citenamefont {Oda},
  \citenamefont {Uno}, \citenamefont {Durrani},\ and\ \citenamefont
  {Milne}}]{Rafiq2006hci}%
  \BibitemOpen
  \bibfield  {author} {\bibinfo {author} {\bibfnamefont {M.~A.}\ \bibnamefont
  {Rafiq}}, \bibinfo {author} {\bibfnamefont {Y.}~\bibnamefont {Tsuchiya}},
  \bibinfo {author} {\bibfnamefont {H.}~\bibnamefont {Mizuta}}, \bibinfo
  {author} {\bibfnamefont {S.}~\bibnamefont {Oda}}, \bibinfo {author}
  {\bibfnamefont {S.}~\bibnamefont {Uno}}, \bibinfo {author} {\bibfnamefont
  {Z.~A.~K.}\ \bibnamefont {Durrani}}, \ and\ \bibinfo {author} {\bibfnamefont
  {W.~I.}\ \bibnamefont {Milne}},\ }\href {\doibase 10.1063/1.2209808}
  {\bibfield  {journal} {\bibinfo  {journal} {Journal of Applied Physics}\
  }\textbf {\bibinfo {volume} {100}},\ \bibinfo {eid} {014303} (\bibinfo {year}
  {2006})}\BibitemShut {NoStop}%
\bibitem [{\citenamefont {Moreira}\ \emph {et~al.}(2011)\citenamefont
  {Moreira}, \citenamefont {Yu}, \citenamefont {Nadal}, \citenamefont
  {Bresson}, \citenamefont {Rosticher}, \citenamefont {Lequeux}, \citenamefont
  {Zimmers},\ and\ \citenamefont {Aubin}}]{Moreira2011ect}%
  \BibitemOpen
  \bibfield  {author} {\bibinfo {author} {\bibfnamefont {H.}~\bibnamefont
  {Moreira}}, \bibinfo {author} {\bibfnamefont {Q.}~\bibnamefont {Yu}},
  \bibinfo {author} {\bibfnamefont {B.}~\bibnamefont {Nadal}}, \bibinfo
  {author} {\bibfnamefont {B.}~\bibnamefont {Bresson}}, \bibinfo {author}
  {\bibfnamefont {M.}~\bibnamefont {Rosticher}}, \bibinfo {author}
  {\bibfnamefont {N.}~\bibnamefont {Lequeux}}, \bibinfo {author} {\bibfnamefont
  {A.}~\bibnamefont {Zimmers}}, \ and\ \bibinfo {author} {\bibfnamefont
  {H.}~\bibnamefont {Aubin}},\ }\href {\doibase 10.1103/PhysRevLett.107.176803}
  {\bibfield  {journal} {\bibinfo  {journal} {Phys. Rev. Lett.}\ }\textbf
  {\bibinfo {volume} {107}},\ \bibinfo {pages} {176803} (\bibinfo {year}
  {2011})}\BibitemShut {NoStop}%
\bibitem [{\citenamefont {Yu}\ \emph {et~al.}(2003)\citenamefont {Yu},
  \citenamefont {Wang},\ and\ \citenamefont {Guyot-Sionnest}}]{Yu2003ncc}%
  \BibitemOpen
  \bibfield  {author} {\bibinfo {author} {\bibfnamefont {D.}~\bibnamefont
  {Yu}}, \bibinfo {author} {\bibfnamefont {C.}~\bibnamefont {Wang}}, \ and\
  \bibinfo {author} {\bibfnamefont {P.}~\bibnamefont {Guyot-Sionnest}},\ }\href
  {\doibase 10.1126/science.1084424} {\bibfield  {journal} {\bibinfo  {journal}
  {Science}\ }\textbf {\bibinfo {volume} {300}},\ \bibinfo {pages} {1277}
  (\bibinfo {year} {2003})}\BibitemShut {NoStop}%
\bibitem [{\citenamefont {Talapin}\ \emph {et~al.}(2010)\citenamefont
  {Talapin}, \citenamefont {Lee}, \citenamefont {Kovalenko},\ and\
  \citenamefont {Shevchenko}}]{Talapin2010poc}%
  \BibitemOpen
  \bibfield  {author} {\bibinfo {author} {\bibfnamefont {D.~V.}\ \bibnamefont
  {Talapin}}, \bibinfo {author} {\bibfnamefont {J.-S.}\ \bibnamefont {Lee}},
  \bibinfo {author} {\bibfnamefont {M.~V.}\ \bibnamefont {Kovalenko}}, \ and\
  \bibinfo {author} {\bibfnamefont {E.~V.}\ \bibnamefont {Shevchenko}},\ }\href
  {\doibase 10.1021/cr900137k} {\bibfield  {journal} {\bibinfo  {journal}
  {Chemical Reviews}\ }\textbf {\bibinfo {volume} {110}},\ \bibinfo {pages}
  {389} (\bibinfo {year} {2010})},\ \bibinfo {note} {pMID:
  19958036}\BibitemShut {NoStop}%
\bibitem [{\citenamefont {Norris}\ \emph {et~al.}(2008)\citenamefont {Norris},
  \citenamefont {Efros},\ and\ \citenamefont {Erwin}}]{Norris2008dn}%
  \BibitemOpen
  \bibfield  {author} {\bibinfo {author} {\bibfnamefont {D.~J.}\ \bibnamefont
  {Norris}}, \bibinfo {author} {\bibfnamefont {A.~L.}\ \bibnamefont {Efros}}, \
  and\ \bibinfo {author} {\bibfnamefont {S.~C.}\ \bibnamefont {Erwin}},\ }\href
  {\doibase 10.1126/science.1143802} {\bibfield  {journal} {\bibinfo  {journal}
  {Science}\ }\textbf {\bibinfo {volume} {319}},\ \bibinfo {pages} {1776}
  (\bibinfo {year} {2008})}\BibitemShut {NoStop}%
\bibitem [{\citenamefont {Liu}\ \emph {et~al.}(2010)\citenamefont {Liu},
  \citenamefont {Pourret},\ and\ \citenamefont {Guyot-Sionnest}}]{Liu2010mae}%
  \BibitemOpen
  \bibfield  {author} {\bibinfo {author} {\bibfnamefont {H.}~\bibnamefont
  {Liu}}, \bibinfo {author} {\bibfnamefont {A.}~\bibnamefont {Pourret}}, \ and\
  \bibinfo {author} {\bibfnamefont {P.}~\bibnamefont {Guyot-Sionnest}},\ }\href
  {\doibase 10.1021/nn101376u} {\bibfield  {journal} {\bibinfo  {journal} {ACS
  Nano}\ }\textbf {\bibinfo {volume} {4}},\ \bibinfo {pages} {5211} (\bibinfo
  {year} {2010})}\BibitemShut {NoStop}%
\bibitem [{\citenamefont {Mott}(1972)}]{Mott1972itc}%
  \BibitemOpen
  \bibfield  {author} {\bibinfo {author} {\bibfnamefont {N.}~\bibnamefont
  {Mott}},\ }\href {\doibase 10.1016/0022-3093(72)90112-3} {\bibfield
  {journal} {\bibinfo  {journal} {Journal of Non-Crystalline Solids}\ }\textbf
  {\bibinfo {volume} {8–10}},\ \bibinfo {pages} {1 } (\bibinfo {year}
  {1972})}\BibitemShut {NoStop}%
\bibitem [{\citenamefont {Kagan}(2012)}]{Kagan2012ros}%
  \BibitemOpen
  \bibfield  {author} {\bibinfo {author} {\bibfnamefont {C.}~\bibnamefont
  {Kagan}},\ }\href {http://meetings.aps.org/link/BAPS.2012.MAR.L18.1}
  {\enquote {\bibinfo {title} {The role of surface ligands in electronic charge
  transport in semiconductor nanocrystal arrays},}\ } (\bibinfo {year}
  {2012}),\ \bibinfo {note} {{APS} March Meeting}\BibitemShut {NoStop}%
\bibitem [{\citenamefont {Zhang}\ and\ \citenamefont
  {Shklovskii}(2004)}]{Zhang2004dos}%
  \BibitemOpen
  \bibfield  {author} {\bibinfo {author} {\bibfnamefont {J.}~\bibnamefont
  {Zhang}}\ and\ \bibinfo {author} {\bibfnamefont {B.~I.}\ \bibnamefont
  {Shklovskii}},\ }\href {\doibase 10.1103/PhysRevB.70.115317} {\bibfield
  {journal} {\bibinfo  {journal} {Phys. Rev. B}\ }\textbf {\bibinfo {volume}
  {70}},\ \bibinfo {pages} {115317} (\bibinfo {year} {2004})}\BibitemShut
  {NoStop}%
\bibitem [{\citenamefont {Mott}(1968)}]{Mott1968cig}%
  \BibitemOpen
  \bibfield  {author} {\bibinfo {author} {\bibfnamefont {N.}~\bibnamefont
  {Mott}},\ }\href {\doibase 10.1016/0022-3093(68)90002-1} {\bibfield
  {journal} {\bibinfo  {journal} {Journal of Non-Crystalline Solids}\ }\textbf
  {\bibinfo {volume} {1}},\ \bibinfo {pages} {1 } (\bibinfo {year}
  {1968})}\BibitemShut {NoStop}%
\bibitem [{\citenamefont {Efros}\ and\ \citenamefont
  {Shklovskii}(1975)}]{Efros1975cga}%
  \BibitemOpen
  \bibfield  {author} {\bibinfo {author} {\bibfnamefont {A.~L.}\ \bibnamefont
  {Efros}}\ and\ \bibinfo {author} {\bibfnamefont {B.~I.}\ \bibnamefont
  {Shklovskii}},\ }\href {\doibase 10.1088/0022-3719/8/4/003} {\bibfield
  {journal} {\bibinfo  {journal} {J. Phys. C: Solid State Phys.}\ }\textbf
  {\bibinfo {volume} {8}},\ \bibinfo {pages} {L49} (\bibinfo {year}
  {1975})}\BibitemShut {NoStop}%
\bibitem [{\citenamefont {Romero}\ and\ \citenamefont
  {Drndic}(2005)}]{Romero2005cba}%
  \BibitemOpen
  \bibfield  {author} {\bibinfo {author} {\bibfnamefont {H.~E.}\ \bibnamefont
  {Romero}}\ and\ \bibinfo {author} {\bibfnamefont {M.}~\bibnamefont
  {Drndic}},\ }\href {\doibase 10.1103/PhysRevLett.95.156801} {\bibfield
  {journal} {\bibinfo  {journal} {Phys. Rev. Lett.}\ }\textbf {\bibinfo
  {volume} {95}},\ \bibinfo {pages} {156801} (\bibinfo {year}
  {2005})}\BibitemShut {NoStop}%
\bibitem [{\citenamefont {Wienkes}\ \emph {et~al.}(2012)\citenamefont
  {Wienkes}, \citenamefont {Blackwell},\ and\ \citenamefont
  {Kakalios}}]{Wienkes2012eti}%
  \BibitemOpen
  \bibfield  {author} {\bibinfo {author} {\bibfnamefont {L.~R.}\ \bibnamefont
  {Wienkes}}, \bibinfo {author} {\bibfnamefont {C.}~\bibnamefont {Blackwell}},
  \ and\ \bibinfo {author} {\bibfnamefont {J.}~\bibnamefont {Kakalios}},\
  }\href {\doibase 10.1063/1.3685491} {\bibfield  {journal} {\bibinfo
  {journal} {Applied Physics Letters}\ }\textbf {\bibinfo {volume} {100}},\
  \bibinfo {eid} {072105} (\bibinfo {year} {2012})}\BibitemShut {NoStop}%
\bibitem [{\citenamefont {Ekimov}\ \emph {et~al.}(1990)\citenamefont {Ekimov},
  \citenamefont {Kudryavtsev}, \citenamefont {Ivanov},\ and\ \citenamefont
  {Efros}}]{Ekimov1990sad}%
  \BibitemOpen
  \bibfield  {author} {\bibinfo {author} {\bibfnamefont {A.}~\bibnamefont
  {Ekimov}}, \bibinfo {author} {\bibfnamefont {I.}~\bibnamefont {Kudryavtsev}},
  \bibinfo {author} {\bibfnamefont {M.}~\bibnamefont {Ivanov}}, \ and\ \bibinfo
  {author} {\bibfnamefont {A.}~\bibnamefont {Efros}},\ }\href {\doibase
  10.1016/0022-2313(90)90010-9} {\bibfield  {journal} {\bibinfo  {journal}
  {Journal of Luminescence}\ }\textbf {\bibinfo {volume} {46}},\ \bibinfo
  {pages} {83 } (\bibinfo {year} {1990})}\BibitemShut {NoStop}%
\bibitem [{\citenamefont {Efros}\ and\ \citenamefont
  {Rosen}(2000)}]{Efros2000eso}%
  \BibitemOpen
  \bibfield  {author} {\bibinfo {author} {\bibfnamefont {A.~L.}\ \bibnamefont
  {Efros}}\ and\ \bibinfo {author} {\bibfnamefont {M.}~\bibnamefont {Rosen}},\
  }\href {\doibase 10.1146/annurev.matsci.30.1.475} {\bibfield  {journal}
  {\bibinfo  {journal} {Annual Review of Materials Science}\ }\textbf {\bibinfo
  {volume} {30}},\ \bibinfo {pages} {475} (\bibinfo {year} {2000})}\BibitemShut
  {NoStop}%
\bibitem [{\citenamefont {Meir}\ \emph {et~al.}(1991)\citenamefont {Meir},
  \citenamefont {Wingreen},\ and\ \citenamefont {Lee}}]{Meir1991tts}%
  \BibitemOpen
  \bibfield  {author} {\bibinfo {author} {\bibfnamefont {Y.}~\bibnamefont
  {Meir}}, \bibinfo {author} {\bibfnamefont {N.~S.}\ \bibnamefont {Wingreen}},
  \ and\ \bibinfo {author} {\bibfnamefont {P.~A.}\ \bibnamefont {Lee}},\ }\href
  {\doibase 10.1103/PhysRevLett.66.3048} {\bibfield  {journal} {\bibinfo
  {journal} {Phys. Rev. Lett.}\ }\textbf {\bibinfo {volume} {66}},\ \bibinfo
  {pages} {3048} (\bibinfo {year} {1991})}\BibitemShut {NoStop}%
\bibitem [{\citenamefont {Maxwell}(1891)}]{Maxwell1891tea}%
  \BibitemOpen
  \bibfield  {author} {\bibinfo {author} {\bibfnamefont {J.~C.}\ \bibnamefont
  {Maxwell}},\ }\href@noop {} {\emph {\bibinfo {title} {A Treatise on
  Electricity and Magnetism}}},\ \bibinfo {edition} {3rd}\ ed.,\ Vol.~\bibinfo
  {volume} {2}\ (\bibinfo  {publisher} {Clarendon},\ \bibinfo {address}
  {Oxford},\ \bibinfo {year} {1891})\ p.~\bibinfo {pages} {57}\BibitemShut
  {NoStop}%
\bibitem [{\citenamefont {Doyle}(1978)}]{Doyle1978cpc}%
  \BibitemOpen
  \bibfield  {author} {\bibinfo {author} {\bibfnamefont {W.~T.}\ \bibnamefont
  {Doyle}},\ }\href {\doibase 10.1063/1.324659} {\bibfield  {journal} {\bibinfo
   {journal} {Journal of Applied Physics}\ }\textbf {\bibinfo {volume} {49}},\
  \bibinfo {pages} {795 } (\bibinfo {year} {1978})}\BibitemShut {NoStop}%
\bibitem [{\citenamefont {Efros}\ and\ \citenamefont
  {Shklovskii}(1984)}]{Efros1984epo}%
  \BibitemOpen
  \bibfield  {author} {\bibinfo {author} {\bibfnamefont {A.~L.}\ \bibnamefont
  {Efros}}\ and\ \bibinfo {author} {\bibfnamefont {B.~I.}\ \bibnamefont
  {Shklovskii}},\ }\href@noop {} {\emph {\bibinfo {title} {Electronic
  Properties of Doped Semiconductors}}}\ (\bibinfo  {publisher}
  {Springer-Verlag},\ \bibinfo {address} {New York},\ \bibinfo {year}
  {1984})\BibitemShut {NoStop}%
\bibitem [{\citenamefont {M\"obius}\ \emph {et~al.}(1992)\citenamefont
  {M\"obius}, \citenamefont {Richter},\ and\ \citenamefont
  {Drittler}}]{Mobius1992cgi}%
  \BibitemOpen
  \bibfield  {author} {\bibinfo {author} {\bibfnamefont {A.}~\bibnamefont
  {M\"obius}}, \bibinfo {author} {\bibfnamefont {M.}~\bibnamefont {Richter}}, \
  and\ \bibinfo {author} {\bibfnamefont {B.}~\bibnamefont {Drittler}},\ }\href
  {\doibase 10.1103/PhysRevB.45.11568} {\bibfield  {journal} {\bibinfo
  {journal} {Phys. Rev. B}\ }\textbf {\bibinfo {volume} {45}},\ \bibinfo
  {pages} {11568} (\bibinfo {year} {1992})}\BibitemShut {NoStop}%
\bibitem [{\citenamefont {Efros}\ \emph {et~al.}(2011)\citenamefont {Efros},
  \citenamefont {Skinner},\ and\ \citenamefont {Shklovskii}}]{Efros2011cgi}%
  \BibitemOpen
  \bibfield  {author} {\bibinfo {author} {\bibfnamefont {A.~L.}\ \bibnamefont
  {Efros}}, \bibinfo {author} {\bibfnamefont {B.}~\bibnamefont {Skinner}}, \
  and\ \bibinfo {author} {\bibfnamefont {B.~I.}\ \bibnamefont {Shklovskii}},\
  }\href {\doibase 10.1103/PhysRevB.84.064204} {\bibfield  {journal} {\bibinfo
  {journal} {Phys. Rev. B}\ }\textbf {\bibinfo {volume} {84}},\ \bibinfo
  {pages} {064204} (\bibinfo {year} {2011})}\BibitemShut {NoStop}%
\bibitem [{\citenamefont {Miller}\ and\ \citenamefont
  {Abrahams}(1960)}]{Miller1960ica}%
  \BibitemOpen
  \bibfield  {author} {\bibinfo {author} {\bibfnamefont {A.}~\bibnamefont
  {Miller}}\ and\ \bibinfo {author} {\bibfnamefont {E.}~\bibnamefont
  {Abrahams}},\ }\href {\doibase 10.1103/PhysRev.120.745} {\bibfield  {journal}
  {\bibinfo  {journal} {Phys. Rev.}\ }\textbf {\bibinfo {volume} {120}},\
  \bibinfo {pages} {745} (\bibinfo {year} {1960})}\BibitemShut {NoStop}%
\bibitem [{\citenamefont {Larkin}\ and\ \citenamefont
  {Khmelnitskii}(1982)}]{Larkin1982aci}%
  \BibitemOpen
  \bibfield  {author} {\bibinfo {author} {\bibfnamefont {A.~I.}\ \bibnamefont
  {Larkin}}\ and\ \bibinfo {author} {\bibfnamefont {D.~E.}\ \bibnamefont
  {Khmelnitskii}},\ }\href@noop {} {\bibfield  {journal} {\bibinfo  {journal}
  {Sov. Phys. JETP}\ }\textbf {\bibinfo {volume} {56}},\ \bibinfo {pages} {647}
  (\bibinfo {year} {1982})}\BibitemShut {NoStop}%
\bibitem [{\citenamefont {Zabrodskii}(1977)}]{Zabrodskii1977}%
  \BibitemOpen
  \bibfield  {author} {\bibinfo {author} {\bibfnamefont {A.~G.}\ \bibnamefont
  {Zabrodskii}},\ }\href@noop {} {\bibfield  {journal} {\bibinfo  {journal}
  {Sov. Phys. Semicond.}\ }\textbf {\bibinfo {volume} {11}},\ \bibinfo {pages}
  {345} (\bibinfo {year} {1977})}\BibitemShut {NoStop}%
\bibitem [{Note1()}]{Note1}%
  \BibitemOpen
  \bibinfo {note} {It can be noted that randomly choosing the donor charge
  $Q_i$ from the interval $[-e/2, +e/2]$ does not result in a system whose
  overall donor charge is neutral for a given realization. One could
  alternatively choose the set $\protect \{Q_i\protect \}$ in such a way that
  the total donor charge is zero, i.e. $\DOTSB \sum@ \slimits@ _i Q_i = 0$. In
  our simulation we find that introducing this constraint provides no
  noticeable effect on either the DOGS or the resistivity.}\BibitemShut {Stop}%
\bibitem [{Note2()}]{Note2}%
  \BibitemOpen
  \bibinfo {note} {Here, following the main ideas of Ref.\ \protect
  \rev@citealp {Zhang2004dos}, we disagree with the estimates in the last
  paragraph of Ref.\ \protect \rev@citealp {Zhang2004dos}'s Sec.\ V. Those
  estimates are appropriate for insulator-coated spheres with distance $d
  \gtrsim D/2$ from each other, but not for much closer touching spheres with
  $d = 2w$. In the latter case, the regions of close contact between
  neighboring spheres play the special role of creating fractionalized donor
  charges, as discussed in Sec.\ \ref {sec:largeoxide}.}\BibitemShut {Stop}%
\bibitem [{\citenamefont {Chen}\ \emph {et~al.}(2011)\citenamefont {Chen},
  \citenamefont {Skinner},\ and\ \citenamefont {Shklovskii}}]{Chen2011cci}%
  \BibitemOpen
  \bibfield  {author} {\bibinfo {author} {\bibfnamefont {T.}~\bibnamefont
  {Chen}}, \bibinfo {author} {\bibfnamefont {B.}~\bibnamefont {Skinner}}, \
  and\ \bibinfo {author} {\bibfnamefont {B.~I.}\ \bibnamefont {Shklovskii}},\
  }\href {\doibase 10.1103/PhysRevB.84.245304} {\bibfield  {journal} {\bibinfo
  {journal} {Phys. Rev. B}\ }\textbf {\bibinfo {volume} {84}},\ \bibinfo
  {pages} {245304} (\bibinfo {year} {2011})}\BibitemShut {NoStop}%
\bibitem [{Note3()}]{Note3}%
  \BibitemOpen
  \bibinfo {note} {In our recent paper \cite {Chen2011cci}, we studied the
  hypothetical case where cations are large enough that only one cation can
  enter a pore in the NC array. In this case, due to the Coulomb interaction,
  the cations form a crystal structure within the pores of the crystalline NC
  array. This situation is different from the model considered in Sec.\ \ref
  {sec:largeoxide}, where ions are small and are introduced at relatively large
  temperature. In Ref.\ \cite {Chen2011cci} we argued that in the former case
  the peak in capacitance corresponding to the 1S shell splits into two
  delta-function-like peaks, such that one electron enters every NC at two
  particular values of the voltage.}\BibitemShut {Stop}%
\end{thebibliography}%
\end{document}